\documentclass[journal, twocolumn, letter]{IEEEtran}

\usepackage{amsmath}
\usepackage{amsfonts}
\usepackage{graphics}
\usepackage{pstricks}
\usepackage{array}
\usepackage{epsf}
\usepackage{graphicx}
\usepackage{rotating}
\usepackage{cite}
\usepackage{pstool}
\usepackage{color}
\usepackage{acronym}
\usepackage{multirow}
\usepackage[font=small]{caption}
\usepackage{subcaption}

\acrodef{fdf}[FDF]{Frequency Divider Formula}
\acrodef{gfdf}[GFDF]{Generalized Frequency Divider Formula}
\acrodef{pmu}[PMU]{Phasor Measurement Unit}
\acrodef{coi}[CoI]{Center of Inertia}
\acrodef{pll}[PLL]{Phase-Locked Loop}
\acrodef{srf}[SRF-PLL]{Synchronous Reference Frame PLL}
\acrodef{dae}[DAE]{Differential-Algebraic Equation}
\acrodef{ode}[ODE]{Ordinary Differential Equation}
\acrodef{rocof}[RoCoF]{Rate of Change of Frequency}
\acrodef{rocop}[RoCoP]{Rate of Change of Power}
\acrodef{irish}[AIITS]{All-Island Irish Transmission System}
\acrodef{cig}[CIG]{Converter-Interfaced Generator}
\acrodef{pfc}[PFC]{Primary Frequency Control}
\acrodef{ffc}[FFR]{Fast Frequency Response}
\acrodef{avr}[AVR]{Automatic Voltage Regulation}
\acrodef{sm}[SM]{synchronous machine}
\acrodef{agc}[AGC]{Automatic Generation Control}
\acrodef{lpf}[LPF]{Low-Pass Filter}
\acrodef{ciess}[CI-ESS]{Converter-Interfaced Energy Storage System}
\acrodef{tcl}[TCL]{Thermostatically-Controlled Load}
\acrodef{pss}[PSS]{Power System Stabiliser}
\acrodef{dfig}[DFIG]{Doubly-Fed Induction Generators}
\acrodef{vsg}[VSG]{Virtual Synchronous Generator}
\acrodef{wpp}[WPP]{Wind Power Plant}
\acrodef{pdf}[PDF]{Probability Density Function}
\acrodef{vdl}[VDL]{Voltage Dependent Load}
\acrodef{der}[DER]{Distributed Energy Resource}
\acrodef{qss}[QSS]{Quasi-Steady-State}

\newcommand{\bfg}[1]{\boldsymbol{#1}}

\newcommand{\bfp}[1]{\bar{\boldsymbol{#1}}}

\newcommand{\bfb}[1]{\boldsymbol{\rm #1}}

\renewcommand{\Re}{{\rm Re}}
\renewcommand{\Im}{{\rm Im}}
\newcommand{\Bhk}{B_{hk}}
\newcommand{\Bhh}{B_{hh}}
\newcommand{\Ghk}{G_{hk}}
\newcommand{\Ghh}{G_{hh}}
\newcommand{\Yhk}{\bar{Y}_{hk}}
\newcommand{\Dt}{\dot}
\newcommand{\dx}[1]{#1_{\rm d}}
\newcommand{\qx}[1]{#1_{\rm q}}
\newcommand{\ds}[1]{#1_{\rm s, d}}
\newcommand{\qs}[1]{#1_{\rm s, q}}
\newcommand{\dr}[1]{#1_{\rm r, d}}
\newcommand{\qr}[1]{#1_{\rm r, q}}
\newcommand{\thk}{\theta_{hk}}
\newcommand{\vh}{v_h}
\newcommand{\vk}{v_k}
\newcommand{\dg}{\delta_{\rm r}}
\newcommand{\wg}{\omega_{\rm r}}
\newcommand{\vg}{\bar{v}_{\rm s}}
\newcommand{\vgdot}{\dot{\bar{v}}_{\rm s}}

\newcommand{\ig}{\bar{\ii}_{\rm s}}
\newcommand{\igdot}{\dot{\bar{\ii}}_{\rm s}}
\newcommand{\vhbar}{\bar{v}_h}
\newcommand{\vhdot}{\dot{v}_h}
\newcommand{\shdot}{\dot{\bar{s}}_h}
\newcommand{\vhbardot}{\dot{\bar{v}}_h}
\newcommand{\etah}{\bar{\eta}_h}
\newcommand{\etak}{\bar{\eta}_k}
\newcommand{\uh}{u_h}
\newcommand{\rhoh}{\varrho_h}
\newcommand{\rhok}{\varrho_k}
\newcommand{\uk}{u_k}

\newcommand{\eh}{\bar{\zeta}_h}
\newcommand{\ek}{\bar{\zeta}_k}
\newcommand{\ph}{p_h}
\newcommand{\qh}{q_h}
\newcommand{\sh}{\bar{s}_h}
\newcommand{\sho}{\bar{s}_{ho}}
\newcommand{\shk}{\bar{s}_{hk}}
\newcommand{\khk}{\bar{\ii}_{hk}}
\renewcommand{\th}{\theta_h}
\newcommand{\wh}{\omega_h}
\newcommand{\wk}{\omega_k}
\newcommand{\tk}{\theta_k}
\renewcommand{\to}{\theta_o}
\newcommand{\tho}{\theta_{h}}
\newcommand{\tko}{\theta_{k}}
\newcommand{\wo}{\omega_o}
\newcommand{\fp}[2]{\frac{\partial #1}{\partial #2}}
\newcommand{\pd}[1]{\tfrac{\partial \ii_{\rm s, d}}{\partial #1}}
\newcommand{\pq}[1]{\tfrac{\partial \ii_{\rm s, q}}{\partial #1}}
\newcommand{\sk}{{\textstyle \sum}_{k=1}^{n}}
\newcommand{\shnk}{{\textstyle \sum}_{h \ne k}^{n}}
\newcommand{\ii}{\imath}
\newcommand{\jj}{\jmath}
\newcommand{\ih}{\bar{\ii}_{h}}
\newcommand{\ihdot}{\dot{\bar{\ii}}_{h}}

\newcommand{\yho}{\bar{Y}_{ho}}
\newcommand{\ytot}{\bar{Y}_{h, \rm tot}}
\newcommand{\yhh}{\bar{Y}_{hh}}
\newcommand{\ivec}{\bfp \ii}
\newcommand{\xdot}{\dot{\bfg x}}
\newcommand{\idot}{\dot{\bar{\bfg \ii}}}
\newcommand{\vvec}{\bfp v}
\newcommand{\vdot}{\dot{\bfp v}}
\newcommand{\phk}{p_{hk}}
\newcommand{\qhk}{q_{hk}}
\newcommand{\svec}{\bar{\bfg s}}
\newcommand{\uvec}{\bfg u}

\newcommand{\sdot}{\dot{\bar{\bfg s}}}
\newcommand{\xih}{\bar{\xi}_h}
\newcommand{\evec}{\bar{\bfg \eta}}
\newcommand{\pdot}{\dot{\bfg p}}
\newcommand{\qdot}{\dot{\bfg q}}
\newcommand{\rhovec}{\bfg \varrho}

\newcommand{\wvec}{\bfg \omega}
\newcommand{\zvec}{\bar{\bfg \zeta}}
\newcommand{\tvec}{\bfg \theta}
\newcommand{\Sbus}{\bar{\bfb S}}
\newcommand{\Ibus}{\bar{\bfb I}}

\newcommand{\Bbus}{\bfb B}
\newcommand{\Gbus}{\bfb G}
\newcommand{\Ybus}{\bar{\bfb Y}}

\graphicspath{{./figs/}}
\DeclareGraphicsExtensions{.eps,.png,.pdf}

\begin{document}
\title{Complex Frequency}
\author{%
  Federico Milano, {\em IEEE Fellow}%
  \thanks{F.~Milano is with the School of Electrical \& Electronic
    Engineering, University College Dublin, Belfield, Ireland.
    E-mail: federico.milano@ucd.ie}%
  \thanks{This work was supported by Science Foundation Ireland, by
    funding F.~Milano under project AMPSAS, Grant No.~SFI/15/IA/3074;
    and by the European Commission by funding F.~Milano under project
    EdgeFLEX, Grant No.~883710.}%
}

\maketitle

\begin{abstract}
  The paper introduces the concept of \textit{complex frequency}.  The
  imaginary part of the complex frequency is the variation with
  respect of a synchronous reference of the local bus frequency as
  commonly defined in power system studies.  The real part is defined
  based on the variation of the voltage magnitude.  The latter term is
  crucial for the correct interpretation and analysis of the variation
  of the frequency at each bus of the network.  The paper also
  develops a set of differential equations that describe the link
  between complex powers and complex frequencies at network buses in
  transient conditions.  No simplifications are assumed except for
   the usual approximations of the models utilized for the
  transient stability analysis of power systems.   A
  variety of analytical and numerical examples show the applications
  and potentials of the proposed concept.
\end{abstract}


\section{Introduction}
\label{sec:intr}

%\subsection{Motivation}
%\label{Motivation}

A well-known and accepted definition of the frequency of a signal
$x(t) = X_m(t) \, \cos(\vartheta(t))$ is given in the IEEE
Std.~IEC/IEEE 60255-118-1 \cite{IEEE118}, as follows:
\begin{equation}
  \label{eq:IEEE118}
  f(t) = \frac{1}{2\pi} \dot{\vartheta}(t) = 
  \frac{1}{2\pi} \dot{\theta}(t) + f_o \, ,
\end{equation}
where $\theta$ is the phase difference, in radians, between the
angular position $\vartheta$, also in radians, of the signal $x(t)$
and the phase due to the reference nominal frequency $f_o$, expressed
in Hz.  If the magnitude $X_m$ of the signal is constant, this
definition is adequate.  However, if $X_m$ changes with time, the
definition of the frequency in \eqref{eq:IEEE118} does not provide a
meaningful way to separate the effects of the variations of
$\vartheta$ and $X_m$.  Thus, the definition of frequency in the most
general conditions is a highly controversial concept that has been
discussed at length in the literature (see the interesting discussion
in \cite{8586583} and the references therein).  

This paper provides a novel interpretation of ``frequency'' as complex
quantity, i.e., composed of a real and an imaginary part.  This
\textit{complex frequency} takes into account the time dependency of
both $\vartheta$ and $X_m$.
% In the following, we refrain from taking any position on the
% ``representationalist'' and ``operationalist'' approaches described
% in \cite{8586583} and refer the reader to the vast literature on
% this topic \cite{7005374}.
The focus is on the theory, modeling, simulation and some application
aspects of the proposed definition.  The proposed complex frequency
allows a neat and compact representation as well as a consistent
interpretation of frequency variations in ac power systems.  The
complex frequency is also capable of explaining the interactions among
active and reactive power injections at buses and flows in network
branches.  It is important to note that the proposed approach does not
attempt to substitute the modeling approaches that go beyond the
classical phasor representation or that focus on analysis of
non-sinusoidal signals (see, for example, \cite{Paolone:2020} for a
state-of-the-art survey on this topic and the several references
therein).  On the contrary, the proposed concept of ``complex
frequency'' is compatible with the approaches that have been proposed
in the literature as it allows interpreting angle and magnitude
variations as complementary components of the same phenomenon,
provided that one accepts to extend the domain of frequency to the
complex numbers.  

This paper focuses on electro-mechanical transients in high-voltage
transmission systems.   Thus, the starting point is
similar to that of \cite{6112197, Divider, 7936594, Maker1}, that is,
the transient conditions during which the magnitude and the phase
angle of bus voltage phasors change according to the inertial response
of synchronous machines and the frequency control of synchronous and
non-synchronous devices.  On the other hand, harmonics, unbalanced
conditions and electro-magnetic transients are not taken into
consideration. 

The resulting formulation is \textit{exact}, in the measure that power
system models based on the $\rm dqo$ transform for voltage and angle
stability analysis are exact;  \textit{general}, as it
provides a framework to study the dynamic effect of any device on the
local frequency variations at network buses; and \textit{systematic},
because it provides with the tools to determine analytically the
impact of each device on bus frequencies.

The remainder of the paper is organized as follows.  Section
\ref{sec:background} provides the background for the proposed
theoretical framework.  Section \ref{sec:derivation} provides the
formal definition of \textit{complex frequency} and its link with
complex power injections, voltages and currents and network topology.
The special cases of constant power and constant current injections as
well as constant impedances are discussed in Section
\ref{sub:special}.  Section \ref{sub:simple} discusses a variety of
relevant approximated expressions that link the complex frequency to
bus power injections.  Section \ref{sec:modeling} illustrates some
applications of the analytical expressions derived in Section
\ref{sec:derivation} to simulation, state estimation and control.
Finally, Section \ref{sec:conclusion} draws conclusions and outlines
future work.

\section{Background}
\label{sec:background}

% This section derives the general expression of the dependency of
% frequency variations at network buses as a function of power
% injections at buses, power flows in the network branches and voltage
% magnitudes.

The starting point is the set of equations that describe the complex
power injections, in per unit, at the $n$ network buses of the system,
say $\svec \in \mathbb{C}^n$, as follows:
\begin{equation}
  \label{eq:st}
  \begin{aligned}
    \svec(t) = \bfg p(t) + \jj \bfg q(t) =
    \bfp v(t) \circ \ivec^*(t) \, ,
  \end{aligned}
\end{equation}
where $\bfg p \in \mathbb{R}^{n \times 1}$ and
$\bfg q \in \mathbb{R}^{n \times 1}$ are the active and reactive power
injections at network buses, respectively;
$\vvec \in \mathbb{C}^{n \times 1}$ and
$\ivec \in \mathbb{C}^{n \times 1}$ are the voltages and current
injections at network buses; $*$ indicates the conjugate of a complex
quantity; and $\circ$ is the Hadamard product, i.e.~the
element-by-element product of two vectors.\footnote{The Hadarmard
  product of two column vectors $\bfg x$ and $\bfg z$ can be also
  written as $\bfg x \circ \bfg z = {\rm diag}(\bfg x) \, \bfg z$,
  where ${\rm diag}(\bfg x)$ is a diagonal matrix whose element
  $(i,i)$ is the $i$-th element of the vector $\bfg x$.}

 In steady-state, balanced conditions, \eqref{eq:st}
expresses the well-known power flow equations.  However, it is
important to note that, in \eqref{eq:st}, all quantities are assumed
to be time dependent.  The elements of the voltage and current vectors
$\bfp v$ and $\bfp \ii$ that appears in \eqref{eq:st}, in fact, are
not to be interpreted as conventional phasors, but as dynamic
quantities, which in some references are called \textit{Park's
  vectors} \cite{Saccomanno:2003, FDF:2020}.
%
%On the other hand, in steady-state, balanced conditions, \eqref{eq:st}
%expresses the well-known power flow equations.
% 
A Park's vector is a complex quantity obtained from the $\rm dq$-axis
components of the well-known $\rm dqo$ transform.   For
example, for the voltage, one has:
\begin{equation}
  \label{eq:vdq}
  \vvec(t) = \bfg v_{\rm d}(t) + \jj \bfg v_{\rm q}(t) \, .
\end{equation}
where the components $v_{{\rm d}, k}$ and $v_{{\rm q}, k}$ of the
$k$-th element of the vector $\vvec$ are calculated as follows:
\begin{equation}
  \label{eq:dqo}
  \begin{bmatrix}
    v_{{\rm d}, k}(t) \\ v_{{\rm q}, k}(t) \\ v_{{\rm o}, k}(t)
  \end{bmatrix} =
  \bfb P(t) 
  \begin{bmatrix}
    v_{{\rm a}, k}(t) \\ v_{{\rm b}, k}(t) \\ v_{{\rm c}, k}(t)
  \end{bmatrix} ,
\end{equation}
where 
\begin{equation}
  \label{eq:park}
  \bfb P(t) = 
  \sqrt{\tfrac{2}{3}}
  \begin{bmatrix}
    \cos(\to(t) ) & \cos(\to'(t)) & \cos(\to''(t)) \\
    \sin(\to(t) ) & \sin(\to'(t)) & \sin(\to''(t)) \\
    \frac{1}{\sqrt{2}} & \frac{1}{\sqrt{2}} & \frac{1}{\sqrt{2}}
  \end{bmatrix} ,
\end{equation}
and $\to$ is the angle between the phase $\rm a$ and the $\rm q$-axis,
with $\dot \to = \omega_o$, and $\to' = \to - \frac {2\pi}{3}$ and
$\to'' = \to + \frac {2\pi}{3}$.   The same
transformation \eqref{eq:dqo} is applied to the $\rm abc$ currents.
Since no assumption is made on the $\rm abc$ quantities, the $\rm d$-
and $\rm q$-axis components of the Park's vectors $\vvec$ and $\ivec$
and, hence,  \eqref{eq:st} can be assumed to be valid in
transient conditions.
%, i.e., for non-sinusoidal $\rm abc$ quantities.
It is important to note that the reactive power is not well defined
for non-sinusoidal signals.\footnote{Several attempts have
  been done to try to overcome this issue.  Among these works, we
  mention the monograph on the ``instantaneous power theory''
  \cite{IPT}.}  However, it is a common assumption, which effectively
underpins the vast majority of studies on the transient stability of
power systems \cite{Sauer:1998}, to approximate the reactive power as
in \eqref{eq:st}.

The $v_{{\rm o}, k}$ is the $\rm o$-axis or \textit{zero} component
and is null for balanced systems.  If the system is not balanced and
the $\rm o$-axis components are not null, then the vector $\bfg p$ in
\eqref{eq:st} does not represent the total active power injections at
network buses as it does not include the term
$\bfg v_{\rm o} \circ \bfg \imath_{\rm o}$.  The hypothesis of
balanced system is not necessary for the developments presented below.
However, since the focus is on high-voltage transmission systems, in
the remainder of this paper, balanced, positive sequence operating
conditions are assumed.

% Since the $\rm o$-axis is stationary, the components along this axis
% do not contribute to frequency variations and is thus ignored in the
% rest of the paper.

% It is important to note that $\vvec$ and $\ivec$ are functions of
% time and are valid for any transient conditions of a three-phase
% balanced ac grid \cite{FDF:2020}.

For the purposes of the developments given below, it is convenient to
rewrite \eqref{eq:vdq} in polar form:
\begin{equation}
  \label{eq:vpolar}
  \vvec(t) = \bfg v(t) \circ \angle \tvec(t) \, ,
\end{equation}
where $\bfg v = |\vvec|$,
$\angle \tvec = \cos(\tvec) + \jj \sin(\tvec)$ and
\begin{equation}
  \label{eq:theta}
  \tvec(t) = \bfg \vartheta(t) - \to(t) \, ,
\end{equation}
namely, $\tvec$ is the vector of bus voltage phase angles referred to
the rotating $\rm dq$-axis reference frame, $\bfg \vartheta$ are the
bus voltage phase angles referred to a  stationary
 reference and $\to = \int_t \wo dt$ is the angle of the
rotating $\rm dq$-axis reference frame and $\wo$ is the angular
frequency in rad/s of the $\rm dq$-axis reference frame.

From \eqref{eq:IEEE118}, the time derivative of $\tvec$ gives:
\begin{equation}
  \label{eq:w}
  \wvec(t) = \Dt {\tvec}(t) = \Dt {\bfg \vartheta}(t) - \wo(t) \, ,
\end{equation}
where $\wvec$ is the vector of frequency deviations with respect to
the reference frequency at the network buses.
In \cite{IEEE118}, it is assumed that $\wo = 2\pi f_o$ is constant and
equal to the nominal angular frequency of the grid, e.g.,
$\wo = 2\pi \, 60$ rad/s in North American transmission grids.
 Note that $\wo$ being constant is not a requirement of
the transform in \eqref{eq:park}.
However, to carry out the derivations presented in Section
\ref{sec:derivation}, $\wo$ is assumed to be constant when it is
utilized to calculate the values of reactances and susceptances.  As
for the reactive power, this is again a widely-accepted approximation
utilized in RMS models for angle and voltage stability analysis and
consists in  assuming that the link between current
injections and voltages is given by
\begin{equation}
  \label{eq:i}
  \ivec(t) \approx \Ybus \, \vvec(t) \, ,
\end{equation}
where $\Ybus = \Gbus + \jj \Bbus \in \mathbb{C}^{n \times n}$ is the
well-known admittance matrix of the network.  It is important not to
confuse \eqref{eq:i} with the conventional relationship between
current and voltage phasors (in which case \eqref{eq:i} is an exact
equality).  $\ivec$ and $\vvec$ are  Park's vectors, i.e.,
complex quantities with time-varying real and imaginary parts
 and, hence, \eqref{eq:i} represents an approximation of
the dynamics of the grid.  In turn, to obtain \eqref{eq:i}, it is
assumed that, for network inductances and capacitances the
relationships between voltages and currents can be approximated with:

\begin{equation}
  \label{eq:approx}
  \begin{aligned}
    \bar v &= L \Dt {\bar \ii} = L (\frac{d}{dt} + \jj \wo ) \bar \ii
    \approx \jj \wo L \bar \ii = \jj X \bar \ii \, , \\
    \bar \ii &= C \Dt {\bar v} = C (\frac{d}{dt} + \jj \wo ) \bar \ii
    \approx \jj \wo C \bar v = \jj B \bar v \, ,    
  \end{aligned}
\end{equation}
where $\frac{d}{dt}$ is the time derivative relative to the Park
rotating frame; $\jj \wo$ is the term due to the rotation of the Park
reference; and  $L$, $C$, $X$, $B$ are the inductance,
capacitance, reactance and susceptance, respectively.  The quantities
in \eqref{eq:approx} are assumed in absolute values.
In turn, the approximation above assumes that electro-magnetic
transients in the elements of the transmission lines and transformers
are \textit{fast} and can be assumed to be in \ac{qss}.  
The approximation \eqref{eq:approx} is applied also to the equations
of the circuits of the devices connected to the grid, e.g., the
equations of the synchronous machine (see \eqref{eq:syn} in Section
\ref{sub:syn}).  The focus of this paper is, in fact, on
the time scales of electro-mechanical and primary frequency and
voltage control transients, which are a few orders of magnitude slower
than electro-magnetic dynamics.

Merging \eqref{eq:st} and \eqref{eq:i} becomes:
\begin{equation}
  \label{eq:s}
  \begin{aligned}
    \svec(t) = \bfp v(t) \circ [ \Ybus \, \bfp v(t)]^* \, .
  \end{aligned}
\end{equation}
These equations resemble the well-known power flow equations except
for the fact that the voltages {are Park's vectors, not
  phasors,} and, hence, the power injections at buses are, in general,
time-varying quantities.

\subsection{Time Derivative of Algebraic Equations}
\label{sub:ydot}

An important aspect of the developments discussed in the next section
is whether \eqref{eq:s} can be differentiated with respect to an
independent variable and, in particular, with respect to time.  With
this aim, observe that \eqref{eq:i} leads to the well-known \ac{qss}
model for power system angle and voltage transient stability analysis,
as follows \cite{Sauer:1998, Milano:2010}: 
\begin{equation}
  \label{eq:dae}
  \begin{aligned}
    \xdot &= \bfg f (\bfg x, \bfg y) \, , \\
    \bfg 0 &= \bfg g (\bfg x, \bfg y) \, ,
  \end{aligned}
\end{equation}
where $\bfg f \in \mathbb{R}^{n_x + n_y} \mapsto \mathbb{R}^{n_x}$ are
the differential equations;
$\bfg g \in \mathbb{R}^{n_x + n_y} \mapsto \mathbb{R}^{n_y}$ are the
algebraic equations, $\bfg x \in \mathcal{X} \subset \mathbb{R}^{n_x}$
are the state variables; and
$\bfg y \in \mathcal{Y} \subset \mathbb{R}^{n_y}$ are the algebraic
variables.  Equations \eqref{eq:dae} are smooth in
$\mathbb{R}^{n_x + n_y}$ except for a finite number of points at which
discrete events occur, such as line outages and faults.  In practice,
these events can be modeled as \textit{if-then} rules that modify the
structure of $\bfg f$ and $\bfg g$.  There exist, of course, more
sophisticated approaches to model events.  These often involve the
definition of additional variables and equations such as the hybrid
automaton described in \cite{1275601}, or the approach based on
Filippov theory discussed in \cite{9415410}.  For simplicity,
however, but without lack of generality, model \eqref{eq:dae} is
considered in the remainder of this work. 

% As said above, the set of \acp{dae} in \eqref{eq:dae} is continuous
% and differentiable except for a finite set of points at which events
% occurs.

The implicit function theorem indicates that, if the Jacobian matrix
$\partial\bfg g / \partial \bfg y$ is not singular, there exists a
function $\bfg \phi$ such that:
\begin{equation}
  \label{eq:ginv}
  \bfg y = \bfg \phi(\bfg x) \, .
\end{equation}
Equation \eqref{eq:ginv} is often utilized to reduce the set of
\acp{dae} in \eqref{eq:dae} into a set of \acp{ode} that depends only
on $\bfg x$ and $\bfg z$.  In this work, however, \eqref{eq:ginv} is
utilized the other way round, i.e., to guarantee that it is possible
to define the time derivative of $\bfg y$ except for the finite number
of points at which the events occur.  This condition leads to:
\begin{equation}
  \label{eq:ydot}
  \begin{aligned}
    \dot{\bfg y} &= \fp{\bfg \phi}{\bfg x} \, \dot{\bfg x} =
    \left (\fp{\bfg g}{\bfg y} \right )^{-1} \fp{\bfg g}{\bfg x} \, \dot{\bfg x} \\
    &= \left (\fp{\bfg g}{\bfg y} \right )^{-1} \fp{\bfg g}{\bfg x}
    \bfg f(\bfg x, \bfg \phi(\bfg x)) \, .
  \end{aligned}
\end{equation} 
The condition \eqref{eq:ydot} implies that the set of \acp{dae} in
\eqref{eq:dae} is assumed to be \textit{index 1} \cite{Ascher:1998},
which is the form of \acp{dae} that describes most physical systems,
including power systems \cite{Milano:2010}.

The voltage magnitudes $\bfg v$ and phase angles $\tvec$ that appears
in \eqref{eq:s}, and hence also the real and imaginary parts of the
complex power $\svec$, are algebraic variables in the conventional
formulation of \ac{qss} models.  Thus, the assumption of index-1
\acp{dae} allows rewriting the current injections at bus and, hence,
the complex power $\svec$ as functions of state and discrete
variables, as well as of the bus voltages $\vvec$, namely
$\svec(\vvec, \bfg x)$,
%
%\begin{equation}
%  \label{eq:sx}
%  \svec(\bfg x, \bfg z) = \vvec(\bfg x, \bfg z) \circ
%  [\Ybus(\bfg z) \, \vvec(\bfg x, \bfg z)]^* \, ,
%\end{equation}
%
which are smooth, except at the points at which the events occur.
Then, the time derivatives of $\svec$ can be computed with the chain
rule as:
\begin{equation}
  \label{eq:sdot:x}
  \sdot = \frac{\partial \svec}{\partial \vvec} \, \vdot +
  \frac{\partial \svec}{\partial \bfg x} \, \xdot \, .
\end{equation}
The next section of this paper elaborates on \eqref{eq:sdot:x} and
deduces an expression that involves the concept of complex frequency.

% Note that the admittance matrix of the system is assumed to be a
% function of $\bfg z$, hence, the discontinuities of $\vvec$ and
% $\svec$ depends not only on the discontinuities of $\bfg \phi$ and
% $\bfg f$, but also on the events the modify the topology of the
% grid.

\section{Derivation}
\label{sec:derivation}

For the sake of the derivation, it is convenient to drop the
dependency on time and rewrite \eqref{eq:s} in an element-wise
notation.  For a network with $n$ buses, one has:
\begin{equation}
  \label{eq:pf}
  \begin{aligned}
    \ph &=
    \vh \sk \vk \, \left [
      \Ghk \, \cos \thk +
      \Bhk \, \sin \thk \right ] \, ,  \\
    \qh &=
    \vh \sk \vk \, \left [
      \Ghk \, \sin \thk -
      \Bhk \, \cos \thk \right ] \, , 
  \end{aligned}
\end{equation}
where $\Ghk$ and $\Bhk$ are the real and imaginary parts of the
element $(h,k)$ of the network admittance matrix,
i.e.~$\Yhk = \Ghk + \jj \Bhk$; $\vh$ and $\vk$ denote the voltage
magnitudes at buses $h$ and $k$, respectively; and $\thk = \th - \tk$,
where $\th$ and $\tk$ are the voltage phase angles at buses $h$ and
$k$, respectively.  Equations \eqref{eq:pf} and all equations with
subindex $h$ in the remainder of this section are valid for
$h=1, 2, \dots, n$.

Equations \eqref{eq:pf} can be equivalently written as:
\begin{equation}
  \label{eq:pf2}
  \begin{aligned}
    \ph = \sk \phk \, , \quad \text{and} \quad
    \qh = \sk \qhk \, , 
  \end{aligned}
\end{equation}
where
\begin{equation}
  \label{eq:pf3}
  \begin{aligned}
    \phk &=
           \vh \vk \, \left [
           \Ghk \, \cos \thk +
           \Bhk \, \sin \thk \right ] \, ,  \\
    \qhk &=
           \vh \vk \, \left [
           \Ghk \, \sin \thk -
           \Bhk \, \cos \thk \right ] \, .
  \end{aligned}
\end{equation}

Differentiating \eqref{eq:pf} and writing the active power injections
as the sum of two components:
\begin{equation}
  \label{eq:p}
  d \ph = 
  \sk \fp{\ph}{\thk} \, d \thk +
  \sk \fp{\ph}{\vk} \, d \vk 
  \equiv d \ph' + d \ph'' \, ,
\end{equation}
In \eqref{eq:p}, $d \ph$ is the total variation of power at bus $h$;
$d \ph'$ is the quota of the active power that depends on bus voltage
phase angle variations; and $d \ph''$ is the quota of active power
that depends on bus voltage magnitude variations.  From \eqref{eq:s}
and using the same procedure that leads to \eqref{eq:p}, one can
define two components also for the reactive power, as follows:
\begin{equation}
  \label{eq:q}
  d \qh = 
  \sk \fp{\qh}{\thk} \, d \thk +
  \sk \fp{\qh}{\vk} \, d \vk \equiv
  d \qh' + d \qh'' \, .
\end{equation}

From \eqref{eq:pf2}, it is relevant to observe that:
\begin{equation}
  \begin{aligned}
    \fp{\ph}{\thk} =-\qhk \, , \quad \text{and} \quad
    \fp{\qh}{\thk} = \phk \, ,
  \end{aligned}
\end{equation}
which leads to rewrite $d \ph'$ and $d \qh'$ as:
\begin{equation}
  \label{eq:dpq1}
  \begin{aligned}
    d \ph' &=-\sk \qhk \, d \thk \, , \\
    d \qh' &= \sk \phk \, d \thk \, ,
  \end{aligned}
\end{equation}
Recalling that $\thk = \th - \tk$, one has:
\begin{equation}
  \label{eq:dpq1c}
  \begin{aligned}
    d \ph' &= - \qh \, d \tho + \sk \qhk \, d \tko \, , \\
    d \qh' &= \ph \, d \tho - \sk \phk \, d \tko \, ,
  \end{aligned}
\end{equation}
where the identities \eqref{eq:pf2} have been used.
%
%\begin{equation}
%  \label{eq:sh}
%  \sh = \sk \shk \, ,
%\end{equation}
%
%and, hence:
%
%\begin{equation}
%  \label{eq:phqh}
%  \ph = \sk \phk \, , \quad \text{and} \quad \qh = \sk \qhk \, .
%\end{equation}

In the same vein, from \eqref{eq:pf2}, \eqref{eq:p} and \eqref{eq:q},
$d \ph''$ and $d \qh''$ can be rewritten as:
\begin{equation}
  \label{eq:dpq2}
  \begin{aligned}
    d \ph'' &= \frac{\ph}{\vh} \, d \vh + \sk \frac{\phk}{\vk} \, d \vk \, , \\
    d \qh'' &= \frac{\qh}{\vh} \, d \vh + \sk \frac{\qhk}{\vk} \, d \vk \, .
  \end{aligned}
\end{equation}
Let us define the quantity:
\begin{equation}
  \label{eq:u}
  \uh \equiv \ln \left (\vh \right ) \, ,
\end{equation}
 where both the logarithm and its argument are
dimensionless.  With this aim, $\vh$ is in per unit, namely it is the
ratio of two voltages (thus without units).   The
differential of \eqref{eq:u} gives:
\begin{equation}
  \label{eq:du}
  d \uh = \frac{d \vh}{\vh} \, .
\end{equation}
Then, \eqref{eq:dpq2} can be rewritten as:
\begin{equation}
  \label{eq:dpq2c}
  \begin{aligned}
    d \ph'' &= \ph \, d \uh + \sk \phk \, d \uk \, , \\
    d \qh'' &= \qh \, d \uh + \sk \qhk \, d \uk \, .
  \end{aligned} 
\end{equation}
Equations \eqref{eq:dpq1c} and \eqref{eq:dpq2c} can be expressed in terms of
complex powers variations:
\begin{equation}
  \label{eq:ds12}
  \begin{aligned}
    d \sh' &= d \ph' + \jj d \qh' = \jj \sh \, d \tho - \jj \sk \shk \, d \tko \, , \\
    d \sh'' &= d \ph'' + \jj d \qh'' = \sh \, d \uh + \sk \shk \, d \uk \, , \\
  \end{aligned}
\end{equation}
and, finally, defining the complex quantity:
\begin{equation}
  \label{eq:e}
  \eh \equiv \uh + \jj \, \th \, , 
\end{equation}
the total complex power variation is given by:
\begin{equation}
  \label{eq:dsh}
  \begin{aligned}
    d \sh &= d \sh' + d \sh'' \\
    &= d \ph' + d \ph'' + \jj (d \qh' + d \qh'') \\
    &= \sh \, d \eh + \sk \shk \, d \ek^* \, .
  \end{aligned}
\end{equation}
Equation \eqref{eq:dsh} can be written in a compact matrix form, as
follows:
\begin{equation}
  \label{eq:svec}
  d \svec = \svec \circ d \zvec + \Sbus \, d \zvec^* \, ,
\end{equation}
where $\Sbus \in \mathbb{C}^{n \times n}$ is a matrix whose $(h,k)$-th
element is $\shk$.

The expression \eqref{eq:svec} has been obtained in general, i.e.,
assuming a differentiation with respect to a generic independent
parameter.  If this parameter is the time $t$, \eqref{eq:svec} can be
rewritten as a set of differential equations:
\begin{equation}
  \label{eq:sdot}
  \boxed{ \sdot - \svec \circ \evec = \Sbus \, \evec^* }
\end{equation}
where
%
%\begin{equation}
%  \sdot = \Dt \svec \, ,
%\end{equation}
%
%and
%
\begin{equation}
  \label{eq:etadot0}
    \evec = \Dt {\zvec} = \Dt {\uvec} + \jj \Dt {\tvec} \, ,
\end{equation}
and recalling the derivative of $\tvec$ given in \eqref{eq:w} and
defining $\rhovec \equiv \dot{\uvec}$, one obtains:
\begin{equation}
  \label{eq:etadot}
  \boxed{ \evec \equiv \rhovec + \jj \, \wvec }
\end{equation}
We define $\evec$ as the vector of \textit{complex frequencies} of the
buses of an AC grid.  Note that both real and imaginary part of
\eqref{eq:etadot} have, in fact, the dimension of s$^{-1}$, as $\uvec$
is dimensionless  and $\wvec$ is expressed in rad/s.  In
\eqref{eq:etadot}, the imaginary part is the usual angular frequency
(relative to the reference $\wo$).  On the other hand, it is more
involved to determine the physical meaning of the real part of
\eqref{eq:etadot}, $\rhovec$.  From the definition of $\uh$, one has
$\vh = {\rm exp}(\uh)$, that is, the magnitude of the voltage is
expressed as a function whose derivative  is equal to the
function itself.  This concept is key in the theory of Lie groups and
algebra, which defines the space of linear transformations of
generalized ``rates of change'' \cite{Stephani:1990}.

% In the following, we show that the physical behavior of the two
% components of $\evec$ are substantially the same as they are
% \textit{dual} variables of the system in a very similar way as the
% active and reactive power components of the complex power $\svec$
% are.

Equation \eqref{eq:sdot} is the sought expression of the relationship
between frequency variations and power flows in an ac grid.  It
contains the information on how power injections of the devices
connected to the grid impact on the frequency at their point of
connection as well as on the rest of the grid.  In \eqref{eq:sdot},
the elements of $\svec$ are the inputs or \textit{boundary conditions}
at network buses and depend on the devices connected to grid, whereas
$\Sbus$ depends only on network quantities. 

Another way to write \eqref{eq:sdot} is by splitting $\sdot$ into its
components $\sdot'$ and $\sdot''$.  According the definitions of
$\svec'$ and $\svec''$, $\sdot'$ does not depend on $\rhovec$, whereas
$\sdot''$ does not depend on $\wvec$, as follows: 
\begin{equation}
  \label{eq:sdot2}
  \begin{aligned}
    \sdot' &= \jj \svec \circ \wvec - \jj \Sbus \, \wvec \, , \\
    \sdot'' &= \svec \circ \rhovec + \Sbus \, \rhovec \, .
  \end{aligned}
\end{equation}
It is important to note that, in general, the expressions of $\sdot'$
and $\sdot''$ are not known \textit{a priori}.  These components,
however, can be determined using \eqref{eq:sdot}, \eqref{eq:sdot2}
and: 
\begin{equation}
  \label{eq:sdot3}
  \sdot = \sdot' + \sdot'' \, .
\end{equation}
With this regard, Section \ref{sub:syn} explains through an example
how to calculate $\wvec$ and $\rhovec$ based on \eqref{eq:sdot}.
 In the following, \eqref{eq:sdot}, \eqref{eq:sdot2} and
\eqref{eq:sdot3} are utilized to discuss relevant special cases.

\subsection{Alternative Expressions}
\label{sub:alternative}

We now derive \eqref{eq:sdot} in an alternative and more compact
formulation as a function of the currents.  First, observe that:
\begin{equation}
  \label{eq:vdot}
  \vdot = \vvec \circ \evec \, .
\end{equation}
The proof of \eqref{eq:vdot} is given in Appendix \ref{app:vdot}.
%
% From \eqref{eq:vdot}, the \textit{complex frequency} can be formally
% defined as the \textit{time-derivative operator of the Park vector
% of the bus voltages}.
Then, recalling \eqref{eq:st}, the time derivative of $\svec$ with
respect to the $\rm dq$-axis reference frame can be written
as:
\begin{equation}
  \label{eq:sdot:alt1}
  \begin{aligned}
    \sdot &= \frac{d}{dt} (\vvec \circ \ivec^*) \\
    &= \vdot \circ \ivec^* + \vvec \circ \idot^* \\
    &= \vvec \circ \evec \circ \ivec^* + \vvec \circ \idot^* \\ 
    &= \svec \circ \evec + \vvec \circ \idot^* \, ,
  \end{aligned}
\end{equation}
 where the commutative property of the Hadamard product
has been used. 
Substituting \eqref{eq:sdot:alt1} into \eqref{eq:sdot}, one obtains:
\begin{equation}
  \label{eq:sdot:alt2}
  \boxed{\vvec \circ \idot^* = \Sbus \, \evec^*}
\end{equation}

Yet another way to write \eqref{eq:sdot} can be obtained by
 differentiating  with respect of time
\eqref{eq:i}, or, which is the same, dividing each row $h$ of $\Sbus$
by the corresponding voltage $\vhbar$ in \eqref{eq:sdot:alt2}.
 Implementing this division and using \eqref{eq:vdot}
 lead to (see also footnote 1):
\begin{equation}
  \label{eq:sdot:alt3}
  \begin{aligned}
    \idot &= \Ybus \, [\vvec \circ \evec]
    = \Ybus \, {\rm diag}(\vvec) \, \evec \, ,
  \end{aligned}
\end{equation}
and, defining $\Ibus = \Ybus \, {\rm diag}(\vvec)$, one obtains:
\begin{equation}
  \label{eq:idot}
  \boxed{\idot = \Ibus \, \evec}
\end{equation}
%
%A relevant byproduct of \eqref{eq:sdot:alt3} is:
%
%\begin{equation}
%  \label{eq:vdot}
%  \boxed{\vdot = \vvec \circ \evec}
%\end{equation}
%
%A proof of \eqref{eq:vdot} is given in the Appendix.

As per \eqref{eq:sdot}, the right-hand sides of \eqref{eq:sdot:alt2}
and \eqref{eq:idot} depend exclusively on network quantities, whereas
the left-hand side is device dependent.  While equivalent, the
relevant feature of \eqref{eq:sdot:alt2} and \eqref{eq:idot} with
respect to \eqref{eq:sdot} is that the complex frequency vector only
appears once.

It is worth noticing that one can also proceed the other way round,
namely obtain \eqref{eq:sdot} from \eqref{eq:idot}.  In fact, taking
the conjugate of \eqref{eq:idot} and multiplying both sides by
$(\vvec \, \circ)$ and using the identity \eqref{eq:sdot:alt1}, one
re-obtains \eqref{eq:sdot}, as expected.

Finally, we note that \eqref{eq:idot} requires less calculations than
\eqref{eq:sdot}.  Thus, in a software tool where currents are modelled
as state variables and, hence, their first time derivatives are
available as a byproduct of the integration of \eqref{eq:dae},
\eqref{eq:idot} can be an efficient alternative to \eqref{eq:sdot} for
the calculation of $\evec$.

% Some examples of this procedure are given in Section
% \ref{sec:modeling}.

% The form in \eqref{eq:idot} is particularly interesting as it allows
% interpreting the elements of the complex frequency vector $\evec$ as
% \textit{potentials} and the elements of the vector $\idot$ as
% \textit{flows}.

% This interpretation is also consistent with the fact that $\wvec$ is
% defined relative to the reference angular frequency $\wo$, and
% $\uvec$ is a function of the bus voltage magnitudes.

\subsection{Special Cases}
\label{sub:special}

Three cases are considered in this section, namely, constant power
injection, constant current injection, and constant admittance load.
These cases illustrate the utilization of the formulas deduced above,
namely \eqref{eq:sdot}, \eqref{eq:sdot:alt2} and \eqref{eq:idot}.

\subsubsection{Constant Power Injection}

We illustrate first an application of \eqref{eq:sdot} for a constant
power injection, say $\sh = \sho$.  From \eqref{eq:sdot3}, the
boundary condition at the $h$-th bus is:
%
%\begin{equation}
%  \label{eq:pconst1}
%  \sh = \sho \, ,
%\end{equation}
%
%hence:
%
\begin{equation}
  \label{eq:pconst2}
  \shdot = 0 \quad \Rightarrow \quad \shdot' = -\shdot'' \, ,
\end{equation}
%
%and, from \eqref{eq:sdot3}:
%
%\begin{equation}
%  \label{eq:pconst3}
%  \shdot' = -\shdot'' \, ,
%\end{equation}
%
and, from \eqref{eq:sdot2}:
\begin{equation}
  \label{eq:pconst4}
  \jj \sho \, \wh - \jj \sk \shk \, \wk =
  - \sho \, \rhoh - \sk \shk \, \rhok \, ,
\end{equation}
and, from \eqref{eq:sdot}:
\begin{equation}
  \label{eq:pconst5}
  - \sho \, \etah = \sk \shk \, \etak^* \, .
\end{equation}

\subsubsection{Constant Admittance Load}
\label{sub:yconst}

This case illustrates an application of \eqref{eq:sdot:alt2}.  For a
constant admittance load, the current consumption at the $h$-th bus
is:
\begin{equation}
  \label{eq:yconst1}
  \ih = - \yho \, \vhbar \, ,
\end{equation}
where the negative sign indicates that the current is drawn from bus
$h$.  From \eqref{eq:sdot:alt2} and \eqref{eq:yconst1} and recalling
that $\vhbardot = \vhbar \, \etah$ (see Appendix \ref{app:vdot}), one
obtains:
\begin{equation}
  \label{eq:yconst2}
  - \yho^* \, \vh^2 \, \etah^* = \sk \shk \, \etak^* \, .
\end{equation}

The same result can be obtained also from \eqref{eq:sdot}.  The
power consumption at bus $h$ can be written as:
\begin{equation}
  \label{eq:yconst3}
  \sh = \vhbar \, \ih^* = - \yho^* \, \vh^2 \, ,
\end{equation}
which indicates that the power consumption $\sh$ in \eqref{eq:yconst3}
does not depend on $\th$.   This means that, according to
\eqref{eq:ds12}, the term $d \sh'=0$ and, from \eqref{eq:sdot3},
$\shdot = \shdot''$ and $\shdot' = 0$.   From the first
equation of \eqref{eq:sdot2}, one has:
\begin{equation}
  \label{eq:yconst4}
  - \yho^* \, \vh^2 \, \wh = \sh \, \wh = \sk \shk \, \wk \, . 
\end{equation}
Then, from the time derivative of \eqref{eq:yconst3} and the second
equation of \eqref{eq:sdot2}, one has:
\begin{equation}
  \label{eq:yconst4b}
  -2 \yho^* \, \vh \, \vhdot =
  -\yho^* \, \vh^2 \, \rhoh + \sk \shk \, \rhok \, .
\end{equation}
Observing that, from \eqref{eq:du}, $\rhoh = \vhdot/\vh$, then:
\begin{equation}
  \vh \, \vhdot = \vh^2 \, \rhoh \, ,
\end{equation}
and, hence, \eqref{eq:yconst4b} can be rewritten as:
\begin{equation}
  \label{eq:yconst5}
  -\yho^* \, \vh^2 \, \rhoh = \sh \, \rhoh = \sk \shk \, \rhok \, ,
\end{equation}
%
%or, equivalently:
%
%\begin{equation}
%  \label{eq:yconst6}
%  \sh \, \rhoh = \sk \shk \, \rhok \, ,
%\end{equation}
%
The expressions \eqref{eq:yconst4} and \eqref{eq:yconst5} have the
same structure and can be merged into \eqref{eq:yconst2}.  Moreover,
in the summations on the right-hand-sides of \eqref{eq:yconst2}, the
term $\bar{s}_{hh}$ is given by:
\begin{equation}
  \bar{s}_{hh} = \bar{Y}_{hh}^* \, \vh^2 \, ,
\end{equation}
and, defining $\ytot = \yho + \yhh$,
one obtains:
\begin{equation}
  \label{eq:yconst7}
  \begin{aligned}
    - \ytot^* \, \vh^2 \, \etah^* = \shnk \shk \, \etak^* \, ,
  \end{aligned}
\end{equation}
which indicates that the two components of the complex frequency,
$\rhoh$ and $\wh$ at the bus $h$ of a constant admittance load are
linear combinations of the $\rhok$ and $\wk$ at the neighboring buses.
This, in turn and as expected, means that constant admittance loads
are \textit{passive} devices and cannot modify the frequency at their
point of connection but rather ``take'' the frequency that is imposed
by the rest of the grid.  This conclusion generalizes the results of
the appendix of \cite{Maker1} that considers the simplified case of an
admittance load connected to the rest of the system through a lossless
line.

\subsubsection{Constant Current Injection}

This last example shows an application of \eqref{eq:idot}.  For a
constant current injection, we have two cases.  If the magnitude and
phase angle of the current are constant, then $\ihdot = 0$ and, from
\eqref{eq:idot}, one obtains immediately:
\begin{equation}
  \label{eq:iconst1}
  0 = \sk \, \khk \, \etak \, ,
\end{equation}
where, $\khk$ is the $(h,k)$ element of $\Ibus$.  Equivalently, from
\eqref{eq:sdot:alt2}, the condition $\ihdot = 0$ leads to:
\begin{equation}
  \label{eq:iconst2}
  0 = \sk \shk \, \etak^* \, .
\end{equation}
On the other hand, it is unlikely that a device is able to impose the
phase angle of its current injection independently from the phase
angle of its bus voltage.  More likely, a device imposes a constant
magnitude and power factor.  In this case, the phase angle of the
current will depend on the phase angle $\th$ of the voltage at bus $h$
 (see Appendix \ref{app:idot}).  This means that,
according to \eqref{eq:ds12}, the term $d \sh''=0$ and, from
\eqref{eq:sdot3}, $\shdot'' = 0$ and $\shdot = \shdot'$.
 This result generalizes the one obtained in
\cite{Freqload}.  From \eqref{eq:idot},
%\eqref{eq:sdot2} and
%\eqref{eq:sdot:alt2},
a current injection with constant magnitude and power factor leads to:
%
%\begin{equation}
%  \label{eq:iconst3}
%  \begin{aligned}
%    \vhbar \, \ihdot^* &= - \jj \sk \shk \, \wk \, , \\
%    0 &= \sk \shk \, \rhok \, ,
%  \end{aligned}
%\end{equation}
%
%or, equivalently, from \eqref{eq:idot}
%
\begin{equation}
  \label{eq:iconst4}
  \begin{aligned}
    \ihdot &= \jj \sk \, \khk \, \wk \, , \\
    0 &= \sk \, \khk \, \rhok \, ,
  \end{aligned}
\end{equation}
and,  since the angular frequency of the current of a
constant power factor load is the same of the angular frequency $\wh$
of the voltage at the load bus (see Appendix \ref{app:idot} and
\eqref{eq:psidot}),  the first equation of
\eqref{eq:iconst4} can be rewritten as:
\begin{equation}
  \label{eq:iconst5}
  \begin{aligned}
    \Re \{ \xih \} =
    \frac{\dot{\ii}_h}{\ii_h} =
    -\jj \left ( \wh - \sk \frac{\khk}{\ih} \, \wk \right ) \, .
  \end{aligned}
\end{equation}

\subsection{Approximated Expressions}
\label{sub:simple}

The mathematical developments carried out so far have assumed no
simplifications except for neglecting the electro-magnetic dynamics of
network branches.  All formulas that have been deduced are thus
accurate in the measure that the effect of electro-magnetic transients
are negligible.  It is, however, relevant to explore whether the
expressions \eqref{eq:sdot}, \eqref{eq:sdot:alt2} and \eqref{eq:idot}
can be approximated while retaining the information on the
relationship between power injections and frequency variations at
network buses.

 Except during faults and some post-fault transients, it
is not uncommon the case for which one can assume  that
$\vh \approx 1$ pu and that bus voltage phase angle differences are
small, hence $\sin (\th-\tk)\approx \th-\tk$ and
$\cos (\th-\tk)\approx 1$.  These assumptions, which, in turn, are the
well-known approximation utilized in the fast decoupled power flow
method \cite{Stott:1974}, lead to:
\begin{equation}
  \label{eq:shk:app}
  \shk \approx \Yhk^* \, .
\end{equation}
Equation \eqref{eq:shk:app} allows simplifying 
\eqref{eq:sdot} as:
\begin{equation}
  \label{eq:sdot:app1}
  \sdot - \svec \circ \evec \approx \Ybus^* \, \evec^* \, ,
\end{equation}
and \eqref{eq:sdot:alt2} and \eqref{eq:idot} as:
\begin{equation}
  \label{eq:idot:app1}
  \idot \approx \Ybus \, \evec \, .
\end{equation}

One can further simplify the expressions above for high-voltage
transmission systems, for which $\Ybus \approx \jj \Bbus$:
\begin{equation}
  \label{eq:sdot:app2}
  \sdot - \svec \circ \evec \approx - \jj \Bbus \, \evec^* \, ,
\end{equation}
and \eqref{eq:sdot:alt2} and \eqref{eq:idot} as:
\begin{equation}
  \label{eq:idot:app2}
  \idot \approx \jj \Bbus \, \evec \, .
\end{equation}
Then, approximating the term
$\svec \circ \evec \approx \bar{\bfb Y}^*_{\rm diag} \evec$, where
$\bar{\bfb Y}_{\rm diag}$ is a matrix obtained using the diagonal
elements of $\Ybus$, and splitting the real and imaginary part of
$\evec$, \eqref{eq:sdot:app2} leads to:
\begin{equation}
  \label{eq:pdot:app1}
  \pdot' \approx \Bbus' \wvec \, ,
\end{equation}
where $\Bhk' = - \Bhk$ and $\Bhh' = \shnk \Bhk$ are the elements of
$\Bbus'$,
%
%\begin{equation*}
%  \Bhk' = - \Bhk \, ,
%  \quad \text{and} \quad
%  \Bhh' = \shnk \Bhk \, ,
%\end{equation*}
%
and
\begin{equation}
  \label{eq:qdot:app1}
  \qdot'' \approx \Bbus'' \rhovec \, ,
\end{equation}
where $\Bhk'' = - \Bhk$ and $\Bhh'' = - 2 \Bhh$ are the elements of
$\Bbus''$.
%
%\begin{equation*}
%  \Bhk'' = - \Bhk \, ,
%  \quad \text{and} \quad
%  \Bhh'' = - 2 \Bhh \, .
%\end{equation*}
%
Equation \eqref{eq:pdot:app1} is the expression deduced in
\cite{Maker1} and that, with due simplifications, leads to the
frequency divider formulas presented in \cite{Divider}.

% It is interesting to note that, in networks for which
% $|\Bbus| > |\Gbus|$, i.e., low-voltage networks,

Considering the resistive parts of the network branches, the following
dual expressions hold:
\begin{equation}
  \label{eq:pdot:app2}
  \begin{aligned}
    \pdot'' &\approx \Gbus'' \rhovec \, , \quad \quad
    \qdot' &\approx \Gbus' \wvec \, ,
  \end{aligned}
\end{equation}
where the elements of $\Gbus'$ and $\Gbus''$ are defined as
$\Ghk' = \Ghk'' = - \Ghk$, $\Ghh' = \shnk \Ghk$, and
$\Ghh'' = -2 \Ghh$.
%
%\begin{align*}
%  \Ghk' = \; &\Ghk'' = - \Ghk \, , \\
%  \Ghh' = \shnk \Ghk \, , \quad &\text{and} \quad
%  \Ghh'' = -2 \Ghh \, .
%\end{align*}
%
Interestingly, from \eqref{eq:pdot:app2}, it descends that, in lossy
networks, the reactive power can be utilized to regulate the
frequency.

Combining together \eqref{eq:pdot:app1}, \eqref{eq:qdot:app1} and
\eqref{eq:pdot:app2} leads to the following approximated expressions:
\begin{equation}
  \label{eq:s1}
  \begin{aligned}
    \sdot' &\approx \jj \Ybus'^* \, \wvec \, , \quad \quad
    \sdot'' &\approx \Ybus'' \, \rhovec \, .
  \end{aligned}
\end{equation}

\section{Examples}
\label{sec:modeling}

The examples presented below apply the theory developed in the
previous section and discuss applications to power system modeling,
state estimation and control.  In particular, three devices are
discussed, namely, the synchronous machine; the voltage dependent
load; and a converter-interfaced generator with frequency and voltage
control capability.  The objective is to illustrate the methodological
approach discussed above and showcase some problems that the proposed
definition of complex frequency and the expression \eqref{eq:sdot}
make possible to solve.  In the following, all simulation results are
obtained using the software tool Dome \cite{Dome}.   With
this regard, note that, in this software tool, $\wo$ is set to be
equal to the frequency of the \ac{coi}, nevertheless, as usually
assumed in transient stability analysis, the values of reactances and
susceptances are kept constant.  

% Of course, using a time-varying $\omega_o$ significantly complicates
% the model of the network, as it requires to take into account the
% elements of the admittance matrix of the grid depends on $\wo$.
% However, $\wo$ being constant is not a requirement of the complex
% frequency definition. 

% These examples serves to illustrate how \eqref{eq:sdot} can be
% implemented and utilized for practical applications.

\subsection{Ratio between $\wvec$ and $\rhovec$}
\label{sub:ratio}

\begin{figure}[htb!]
  \centering
  \begin{subfigure}[t]{\linewidth}
    \resizebox{0.98\linewidth}{!}{\includegraphics{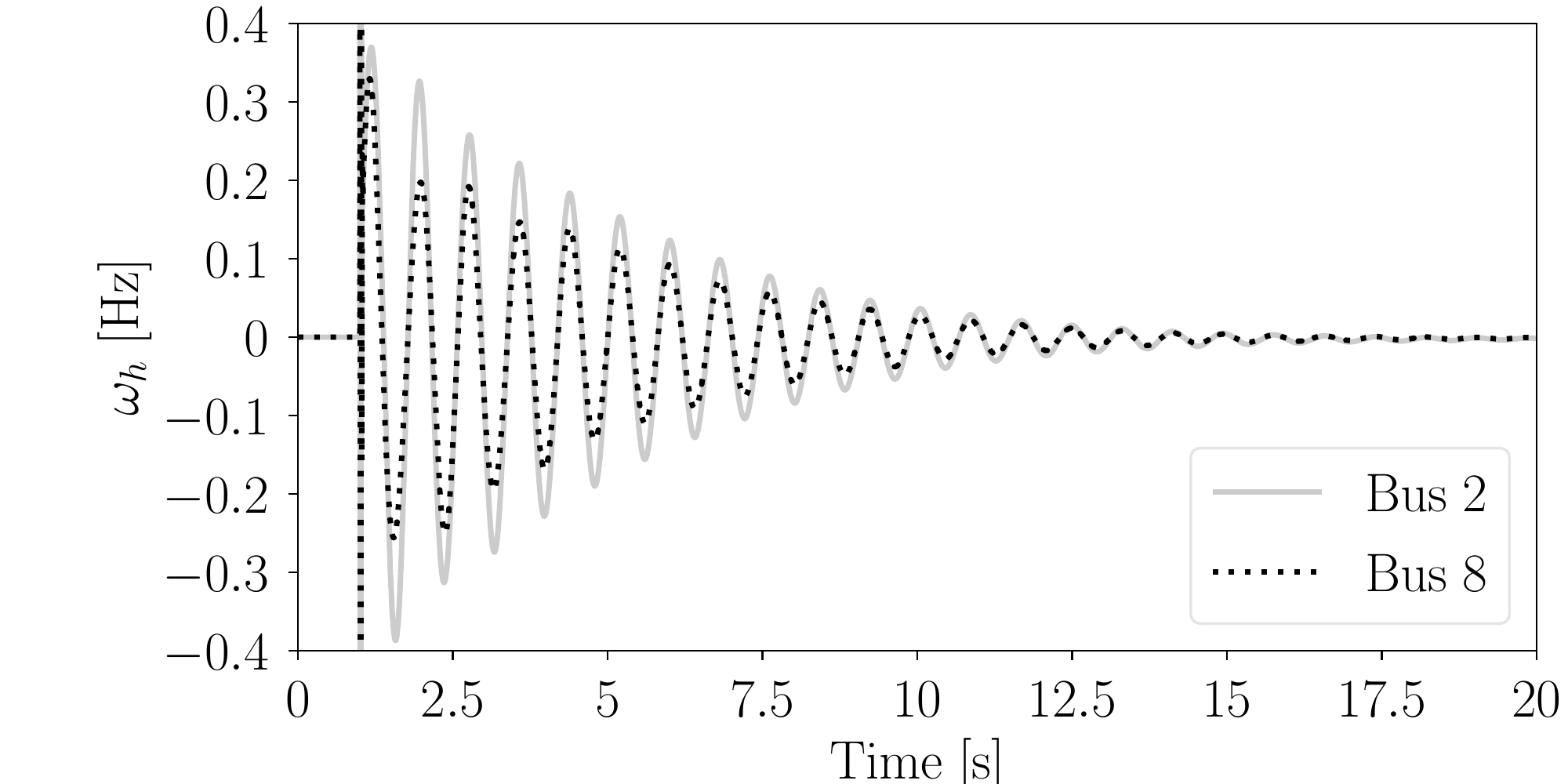}} \\
    \vspace{2mm}
    \resizebox{0.98\linewidth}{!}{\includegraphics{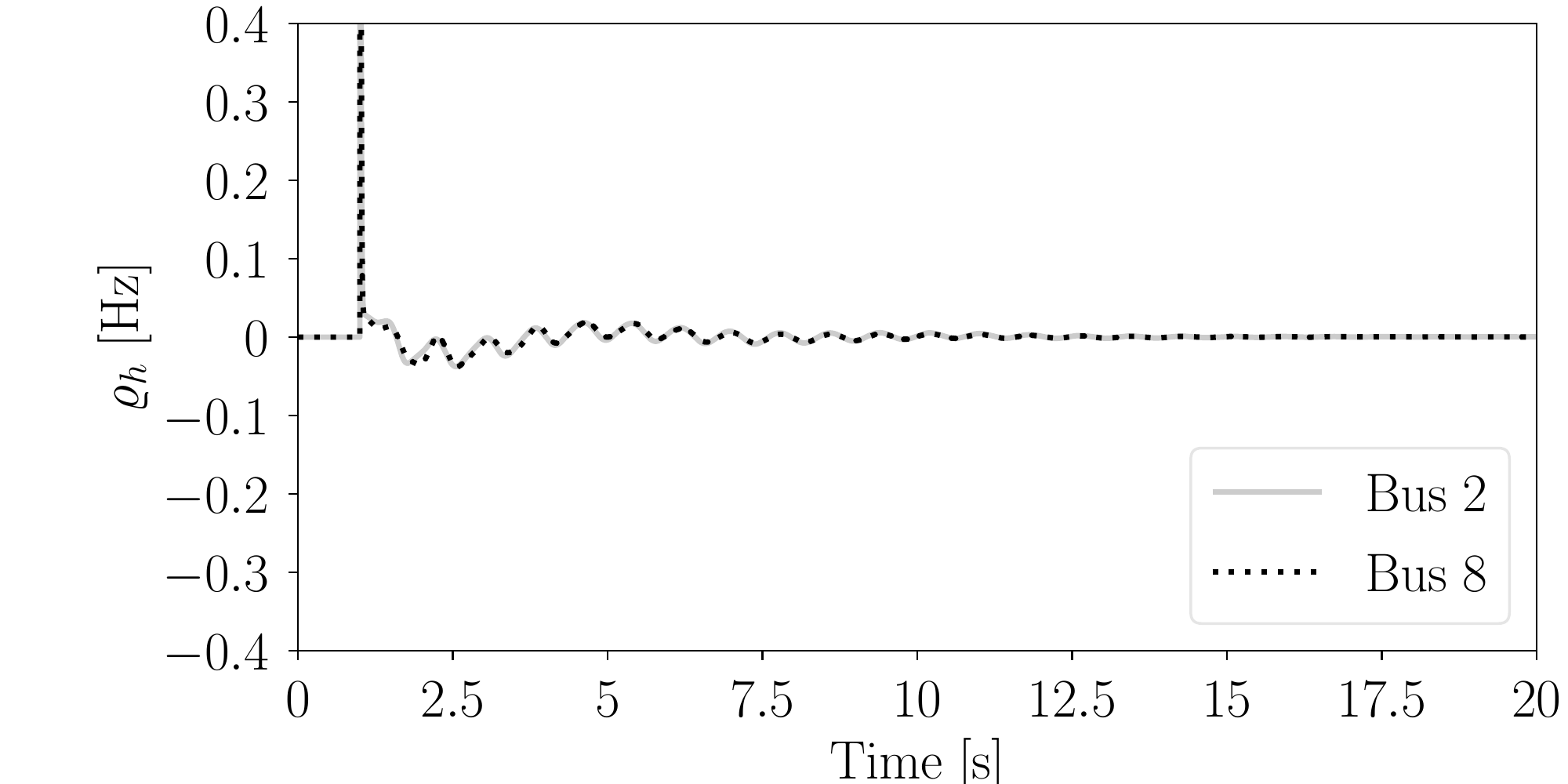}}
    \caption{Outage of the load at bus 5.}
    \label{fig:wscc:ex1}
  \end{subfigure} \\ \vspace{3mm}
  \begin{subfigure}[t]{\linewidth}
    \centering
    \resizebox{0.98\linewidth}{!}{\includegraphics{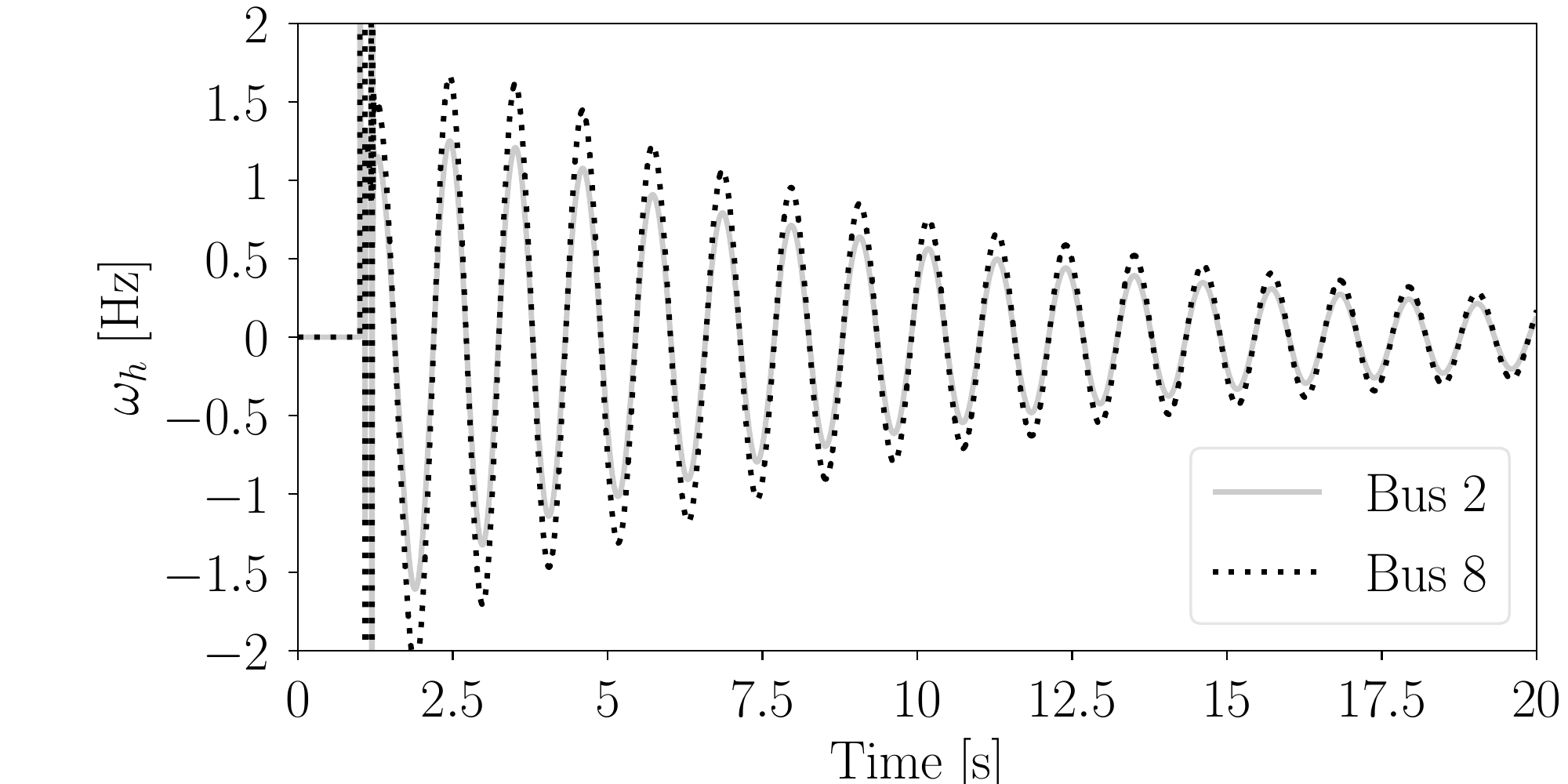}} \\
    \vspace{2mm}
    \resizebox{0.98\linewidth}{!}{\includegraphics{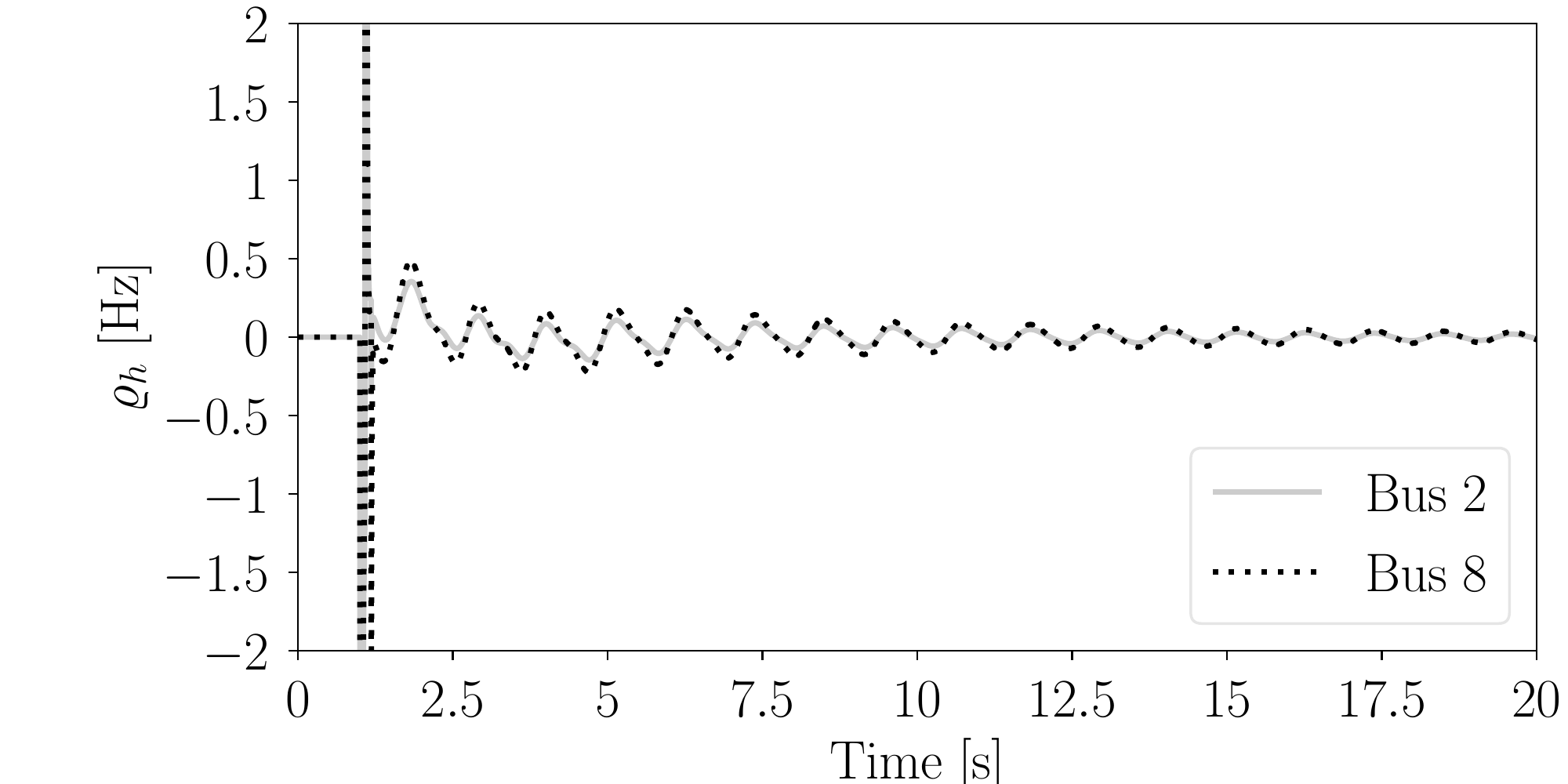}}
    \caption{Fault at bus 7 cleared by tripping line 5-7.}
    \label{fig:wscc:ex2}
  \end{subfigure}
  \caption{Transient behavior of the components of the complex
    frequency for the WSCC 9-bus system.}
  \label{fig:wscc}
\end{figure}

This first example illustrates the transient behavior of the two
components of the complex frequency.  Figure \ref{fig:wscc} shows the
transient behavior of the components of the complex frequency at a
generator and a load bus, bus 2 and bus 8, respectively, for the
well-known WSCC 9-bus system \cite{Sauer:1998}.  The estimation of
$\wh$ and $\rhoh$ at the buses of the grid is obtained through a
synchronous-reference frame \ac{pll} model \cite{Ortega:2018}.
Simulation results show that $|\wh| \gg |\rhoh|$.
%
%\begin{equation}
%  \label{eq:smalludot}
%  |\wvec| \gg |\rhovec| \, ,
%\end{equation}
%
%after a large contingency that triggers frequency variations in the
%system.  The validity of \eqref{eq:smalludot} is, in effect, quite
%general.
This inequality holds for all networks and scenarios that we have
tested for the preparation of this work.

It is important to note that the inequality $|\wh| \gg |\rhoh|$ does
not imply $\shdot' \gg \shdot''$.  As a matter of fact, taking as an
example constant impedance loads, $\shdot' = 0$ and
$\shdot'' = \shdot$.  The rationale behind this observation can be
explained by rewriting \eqref{eq:sdot2} using \eqref{eq:pf2} and an
element-by-element notation:
\begin{equation}
  \label{eq:sdot2bis}
  \begin{aligned}
    \shdot' &= \jj \sk \shk (\wh -\wk) \, , \\
    \shdot'' &= \sk \shk (\rhoh + \rhok ) \, ,
  \end{aligned}
\end{equation}
which indicates that $\shdot'$ is proportional to the
\textit{difference} of the elements of $\wvec$, whereas $\shdot''$ is
proportional to the \textit{sum} of the elements of $\rhovec$.

% Hence, even if the elements of $\rhovec$ are small relatively to
% those of $\wvec$, their effect on the system are not necessarily
% negligible.

\subsection{Implementation of Equation \eqref{eq:sdot}}
\label{sub:syn}

The implementation of \eqref{eq:sdot} in a software tool for the
simulation of power systems can be useful to determine the ``exact''
frequency variations at network buses in a RMS model for transient
stability analysis.  This topic has been discussed and solved under
various hypotheses in \cite{6112197, Divider, 7936594}.  In
particular, the \ac{fdf} proposed in \cite{Divider} is based on
\eqref{eq:pdot:app1}, which is an approximation of \eqref{eq:sdot}.

The expression \eqref{eq:sdot} states the link between the complex
frequency and the rest of the variables of the system.  Thus, in
\eqref{eq:sdot}, the unknowns are $\bar{\bfg \eta}$ or, equivalently,
$\bfg \rho$ and $\bfg \omega$.
%
% Now, it happens that \eqref{eq:sdot} is in the form of a set of
% differential equations in the sense that they depend (in general) on
% the first time derivatives of the states (through
% $\dot{\bar{\bfg s}}$).
%
One can, of course, calculate these quantities by differentiating with
respect to time $\bfg u$ and $\bfg \theta$, respectively.  However,
since $\bfg u$ and $\bfg \theta$ are algebraic variables,
\eqref{eq:dae} does not provide directly the values of their time
derivatives.  So either one has to use some sort of numerical
differentiation (with the issues that arise at discontinuities); use
some sort of estimation (e.g., the PLL utilized in the previous
example); or solve \eqref{eq:sdot}.  The latter is the approach
utilized in this example.

The procedure implemented in the simulations presented in this section
is as follows.
\begin{itemize}
\item The system variables $\bfg x$ and $\bfg y$ are initialized using
  \eqref{eq:dae}.
\item The \acp{dae} defined by \eqref{eq:dae} are integrated using a
  conventional numerical scheme (in this case, the implicit
  trapezoidal method with fixed time step).
\item Discrete events are also handled with a conventional approach.
  Specifically, an approach based on \textit{if-then} rules that
  switch the equations on the right-hand-side of \eqref{eq:dae} has
  been utilized.
\item Finally, \eqref{eq:sdot} -- or, alternatively,
  \eqref{eq:sdot:alt2} or \eqref{eq:idot} -- and the values $\bfg x$,
  $\bfg y$ and $\dot{\bfg x}$ obtained at each step of the time domain
  integration are utilized to calculate $\evec$.\footnote{ The
    procedure described above is an open-loop, i.e., \eqref{eq:sdot}
    are solved ``off-line'' based on the solution obtained by
    integrating \eqref{eq:dae}.  In a closed loop, i.e., if the
    elements of the complex frequency are utilized as inputs to some
    controllers or one wants to take into account the dependence on
    the frequency in the model of a device, then \eqref{eq:dae} has to
    be extended to include \eqref{eq:sdot}.}
\end{itemize}
With regard to the last step, i.e., the determination of $\evec$, note
that \eqref{eq:sdot} can be rewritten as a set of $2n$ real equations,
with unknowns $(\rhovec, \wvec)$.  The right-hand side of
\eqref{eq:sdot} is linear with respect to $\evec$, so it remains to
determine the dependency of the elements of
$\sdot - \svec \circ \evec^*$ -- or, alternatively, of
$\vvec \circ \idot^*$ from \eqref{eq:sdot:alt2} or $\idot$ from
\eqref{eq:idot} -- on $\evec$ and, eventually, on the time derivatives
of the state variables $\xdot$.

We illustrate the procedure using a conventional 4th order model of
the synchronous machine and \eqref{eq:sdot}.  The stator voltage of
the machine with respect to its $\rm dq$-axis reference frame is
linked to the grid voltage $\vh$ with the following equations
\cite{Sauer:1998}:
\begin{equation}
  \label{eq:vg}
  \vg = \ds{v} + \jj \, \qs{v} =
  \vhbar \, \angle (\tfrac{\pi}{2} - \dg) \, ,
\end{equation}
where $\dg$ is th rotor angle of the machine.  The time derivative of
\eqref{eq:vg} gives (see Appendix \ref{app:vdot} for the derivative of
$\vhbar$):
\begin{equation}
  \label{eq:vgdot}
  \vgdot = \vg \, (\etah - \jj \, \wg) \, ,   
\end{equation}
where $\wg$ is the deviation in rad/s of the angular speed of the
machine with respect to $\wo$.   The real and imaginary
terms of \eqref{eq:vgdot} can be written as:
\begin{equation}
  \label{eq:vgdot2}
  \begin{aligned}
    \ds{\dot{v}} &= \ds{v} \rhoh - \qs{v} \wh + \qs{v} \wg \, , \\
    \qs{\dot{v}} &= \qs{v} \rhoh + \ds{v} \wh - \ds{v} \wg \, . \\    
  \end{aligned}
\end{equation}
Then, the stator electrical and magnetic equations are:
\begin{equation}
  \label{eq:syn}
  \begin{aligned}
    \dx{X}' \, \ds{\ii} + R_{\rm a} \, \qs{\ii}  &= - \qs{v} + \qr{e'} \\
    %\lambda_{\rm d}(\bfg x) \, , \\
    %\gamma_{\rm d} \, \qr{e'} + (1 - \gamma_{\rm d}) \, \qr{e''} \, , \\
    R_{\rm a} \, \ds{\ii} - \qx{X}' \, \qs{\ii} &= - \ds{v} + \dr{e'}
    %\lambda_{\rm q}(\bfg x) \, ,
    % \gamma_{\rm q} \, \dr{e'} - (1 - \gamma_{\rm q}) \, \dr{e''} \, ,
  \end{aligned}
\end{equation}
where $\ig = \ds{\ii} + \jj \, \qs{\ii}$ is the stator current.  From
\eqref{eq:syn}, one can thus obtain the expressions of $\ds{\ii}$ and
$\qs{\ii}$ as a function of $\ds{v}$, $\qs{v}$ and the state variables
$\dr{e'}$ and $\qr{e'}$.  In this case, these relationships are
linear, so, one can obtain the current explicitly as:
\begin{equation}
  \label{eq:syn2a}
  \begin{bmatrix}
    \ds{\ii} \\ \qs{\ii}
  \end{bmatrix}
  =
  \begin{bmatrix}
    \dx{X}' & R_{\rm a} \\
    R_{\rm a} & -\qx{X}'
  \end{bmatrix}^{-1}
  \begin{bmatrix}
    - \qs{v} + \qr{e'} \\
    - \ds{v} + \dr{e'}
  \end{bmatrix} \, ,
\end{equation}
or, equivalently:
\begin{equation}
  \label{eq:syn2}
  \begin{aligned}
    \ds{\ii} &= k_{v, \rm dd} \ds{v} + k_{v, \rm dq} \qs{v} +
    k_{e, \rm dd} \dr{e'} + k_{e, \rm dq} \qr{e'} \, , \\
    \qs{\ii} &= k_{v, \rm qd} \ds{v} + k_{v, \rm qq} \qs{v} +
    k_{e, \rm qd} \dr{e'} + k_{e, \rm qq} \qr{e'} \, ,
  \end{aligned}
\end{equation}
where the $k$-coefficients are constant and depend on $R_{\rm a}$,
$\dx{X}'$ and $\qx{X}'$.
Differentiating \eqref{eq:syn2} with respect to time, substituting the
expressions of $\ds{\dot{v}}$ and $\ds{\dot{v}}$ from
\eqref{eq:vgdot2} and applying the chain rule, the time derivatives of
the components of the current can be written in the form:
\begin{equation}
  \label{eq:syn:idot}
  \begin{aligned}
    \ds{\dot{\ii}} &=  \pd{\rhoh} \rhoh + \pd{\wh} \wh + \pd{\wg} \wg + 
    \pd{\dr{e}'} \dr{\dot{e}'} + \pd{\qr{e}'} \qr{\dot{e}'} \, , \\
    %\ds{\dot{\ii}} &= \kd{1} \wh + \kd{2} \rhoh + \kd{3} \wg + 
    %\kd{4} \dr{\dot{e}'} + \kd{5} \qr{\dot{e}'} \, , \\
    \qs{\dot{\ii}} &= \pq{\rhoh} \rhoh + \pq{\wh} \wh +  \pq{\wg} \wg + 
    \pq{\dr{e}'} \dr{\dot{e}'} + \pq{\qr{e}'} \qr{\dot{e}'} \, ,
    %\kq{1} \wh + \kq{2} \rhoh + \kq{3} \wg + 
    %\kq{4} \dr{\dot{e}'} + \kq{5} \qr{\dot{e}'} \, ,
  \end{aligned}
\end{equation}
where, in the right-hand side, $\rhoh$ and $\wh$ are unknown and all
other variables and terms are calculated based on the current solution
of \eqref{eq:dae}.
%
% where the coefficients $\kd{\cdot}$ and $\kq{\cdot}$ depends on the
% machine parameters and/or the terminal bus voltage $\vg$ and
%where the time derivative of $\ds{v}$ and $\qs{v}$ have been
%substituted with the real and imaginary parts of the left-hand side of
%\eqref{eq:vgdot}.
%
Then, from \eqref{eq:sdot:alt2}, one obtains:
\begin{equation}
  \label{eq:syn:sdot}
  \begin{aligned}
    \Re \{ \vhbar \ihdot^* \} &= \Re \{ \vg \igdot^* \} =
    \ds{v} \ds{\dot{\ii}} + \qs{v} \qs{\dot{\ii}} \, , \\
    \Im \{ \vhbar \ihdot^* \} &= \Im \{ \vg \igdot^* \} =
    \qs{v} \ds{\dot{\ii}} - \ds{v} \qs{\dot{\ii}} \, .
  \end{aligned}
\end{equation}
Finally, substituting \eqref{eq:syn:idot} into \eqref{eq:syn:sdot},
one obtains the expressions of the left-hand side of
\eqref{eq:sdot:alt2} at the buses of the synchronous machines.

The very same procedure can be applied to any device of the grid and,
in the vast majority of the cases, the resulting expressions are
linear with respect to $(\rhovec, \wvec)$.  Hence, at each time step
of a time domain simulation one need to solve a problem of the type
$\bfb A \bfg \chi = \bfg b$, where
$\bfg \chi = [\rhovec^T, \wvec^T]^T$, and:
\begin{equation}
  \bfb A =
  \begin{bmatrix}
    [\bfb H + {\rm diag}(\bfg p) - \bfb P_{\rhovec}] &
    \phantom{-}[\bfb K - {\rm diag}(\bfg q) - \bfb P_{\wvec}] \\ 
    [\bfb K + {\rm diag}(\bfg q) - \bfb Q_{\rhovec}] &
    -[\bfb H - {\rm diag}(\bfg p) + \bfb Q_{\wvec}] 
  \end{bmatrix} 
\end{equation}
and
\begin{equation}
  \bfg b =
  \begin{bmatrix}
    \bfb P_{\bfg x} \xdot \\
    \bfb Q_{\bfg x} \xdot
  \end{bmatrix} ,
\end{equation}
where $\bfb H = \Re\{\Sbus\}$ and $\bfb K = \Im\{\Sbus\}$;
$\bfb P_{\rhovec}$, $\bfb P_{\wvec}$, $\bfb Q_{\rhovec}$ and
$\bfb Q_{\wvec}$ are the Jacobian $n \times n$ matrices of $\bfg p$
and $\bfg q$ with respect to $\rhovec$ and $\wvec$, respectively; and
$\bfb P_{\bfg x}$ and $\bfb Q_{\bfg x}$ are the Jacobian
$n \times n_x$ matrices of $\bfg p$ and $\bfg q$ with respect to the
state variables $\bfg x$, respectively.

\begin{figure}[htb!]
  \centering
  \begin{subfigure}[t]{0.98\linewidth}
    \centering
    \resizebox{\linewidth}{!}{\includegraphics{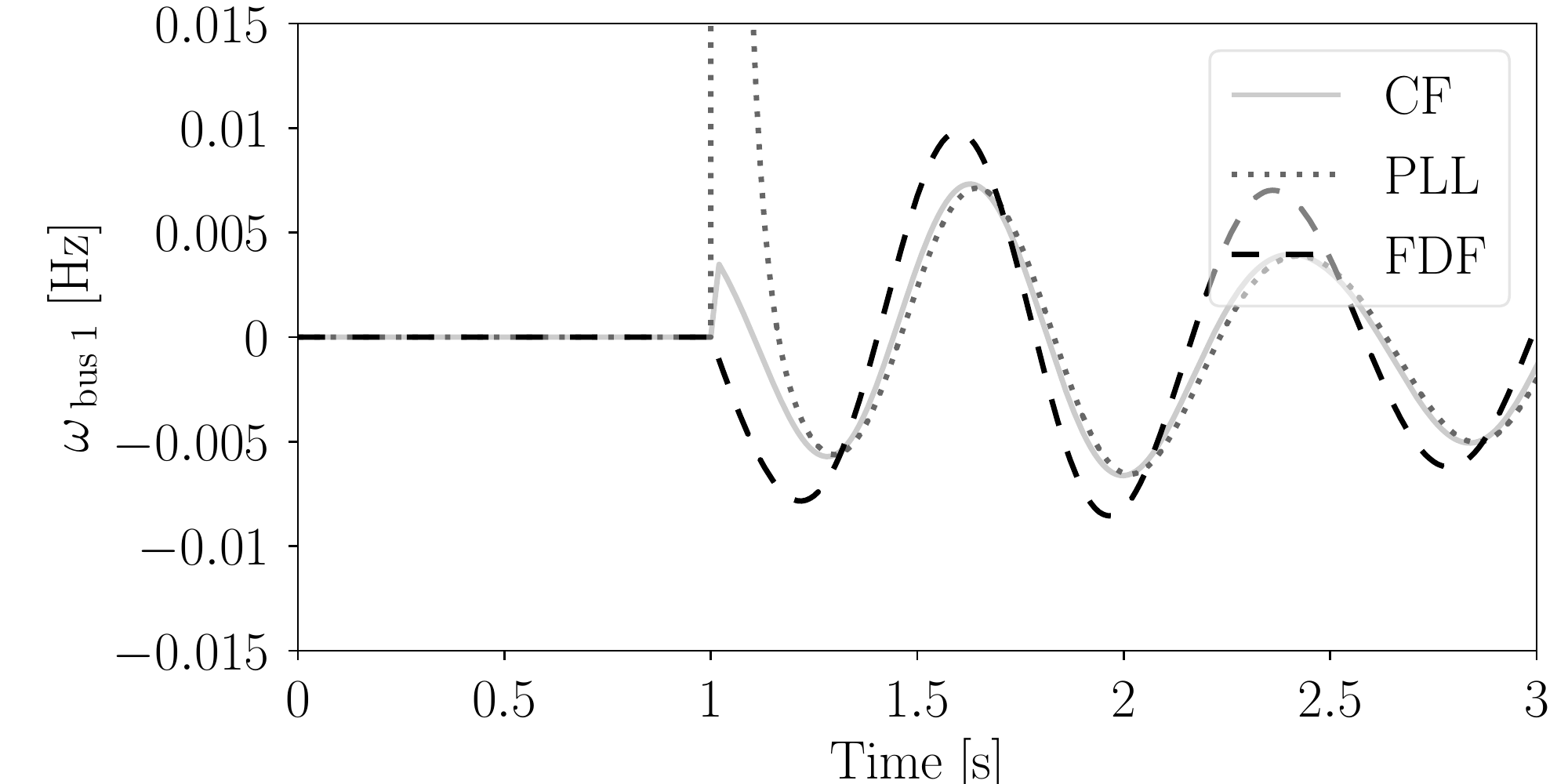}}
    \caption{4th order machine models}
    \label{fig:syn:ex1}
  \end{subfigure} \\ \vspace{2mm}
  \begin{subfigure}[t]{0.98\linewidth}
    \centering
    \resizebox{\linewidth}{!}{\includegraphics{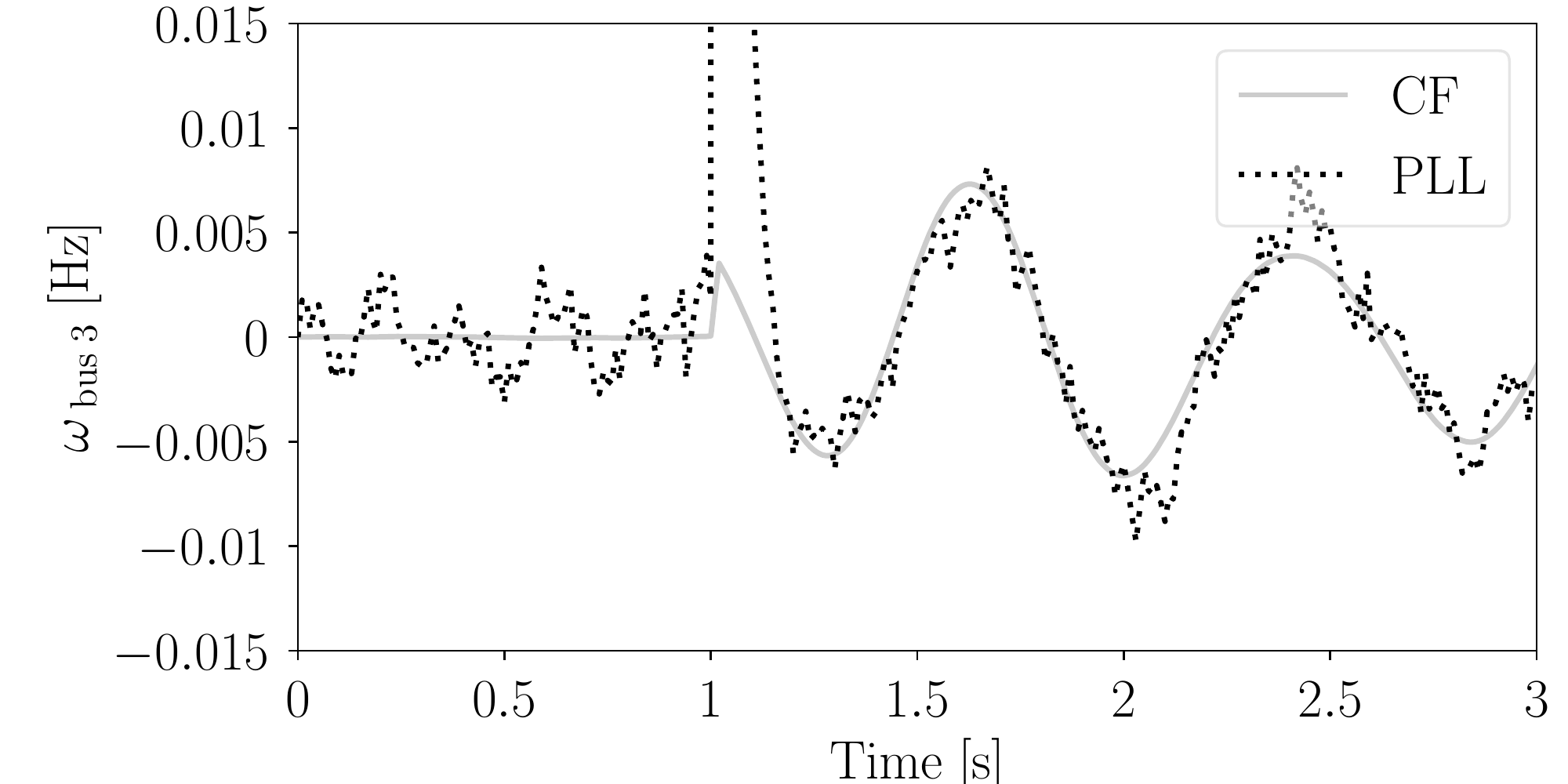}}
    \caption{4th order machine models \& noise}
    \label{fig:syn:ex4}
  \end{subfigure} \\ \vspace{2mm}
  \begin{subfigure}[t]{0.98\linewidth}
    \centering
    \resizebox{\linewidth}{!}{\includegraphics{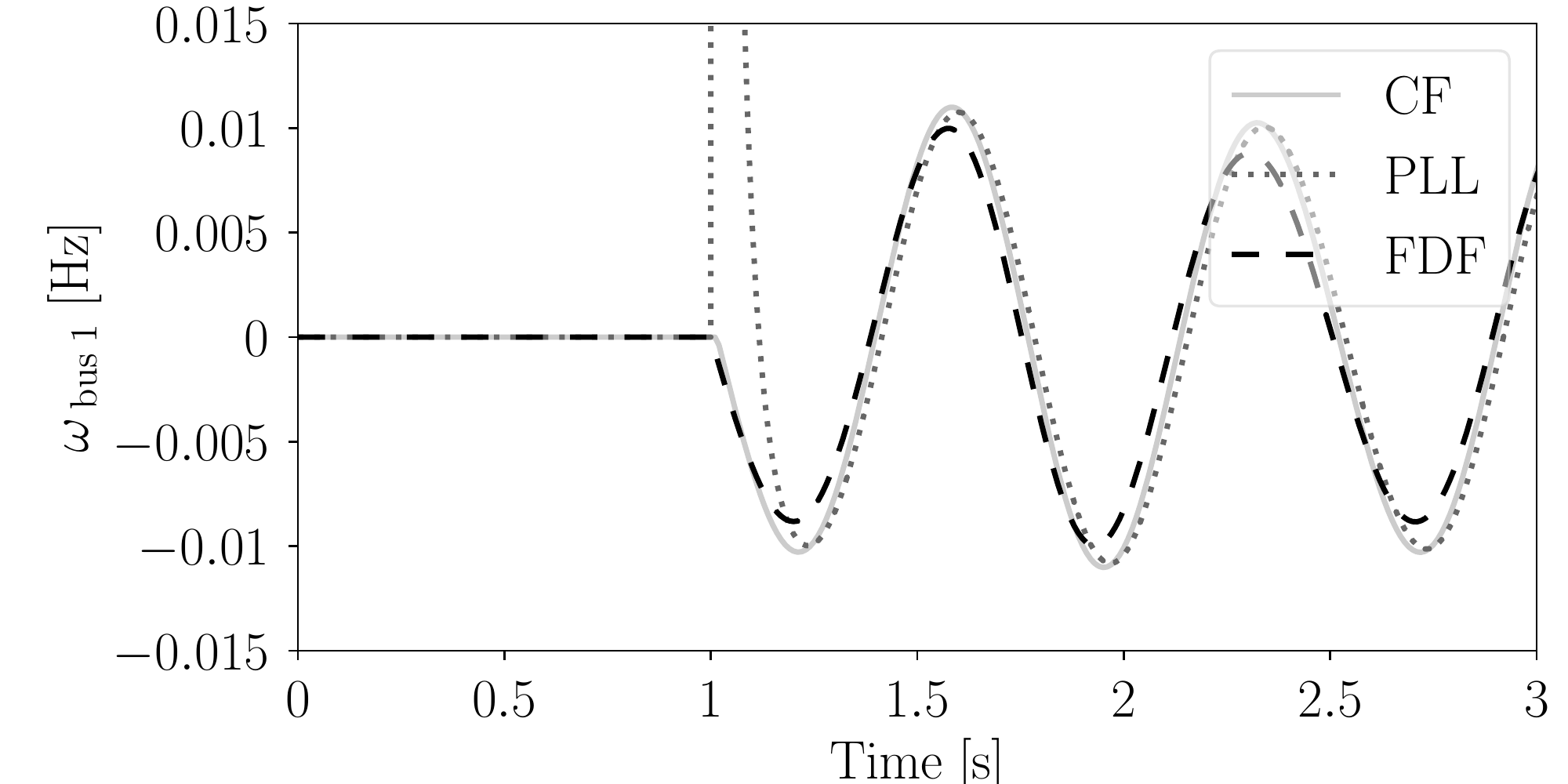}}
    \caption{2nd order machine models}
    \label{fig:syn:ex2}
  \end{subfigure} \\ \vspace{2mm}
  \begin{subfigure}[t]{0.98\linewidth}
    \centering
    \resizebox{\linewidth}{!}{\includegraphics{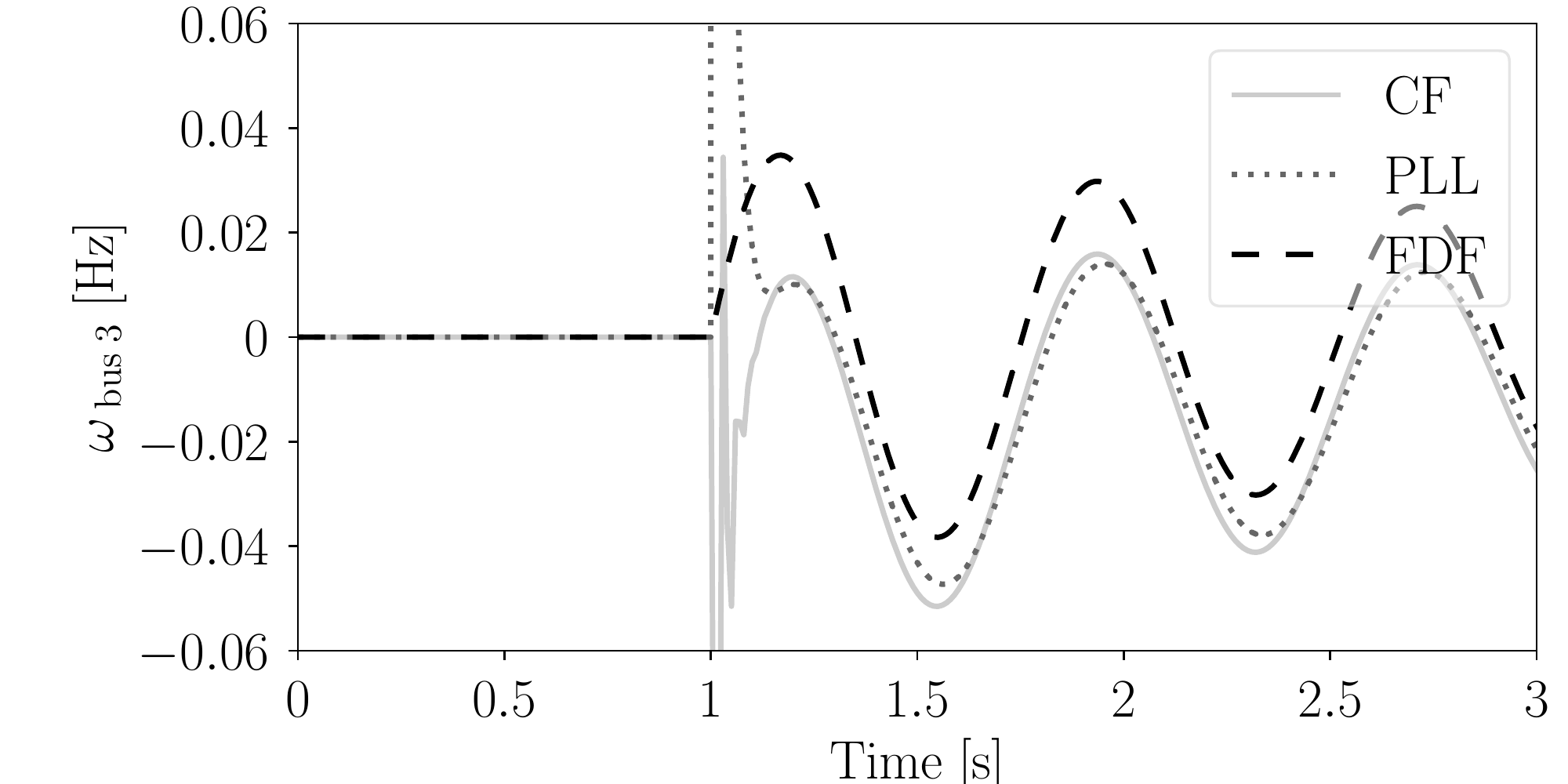}}
    \caption{\ac{cig} connected at bus 3}
    \label{fig:syn:ex3}
  \end{subfigure} \\
  \caption{Comparison of bus frequencies using \ac{pll}, \ac{fdf} and
    the proposed approach based on the complex frequency.  The plots
    refers to the WSCC 9-bus system following the disconnection at
    $t=1$ s of the load at bus 5.}
  \label{fig:syn}
\end{figure}

\begin{figure}[htb!]
  \centering
  \resizebox{0.85\linewidth}{!}{\includegraphics{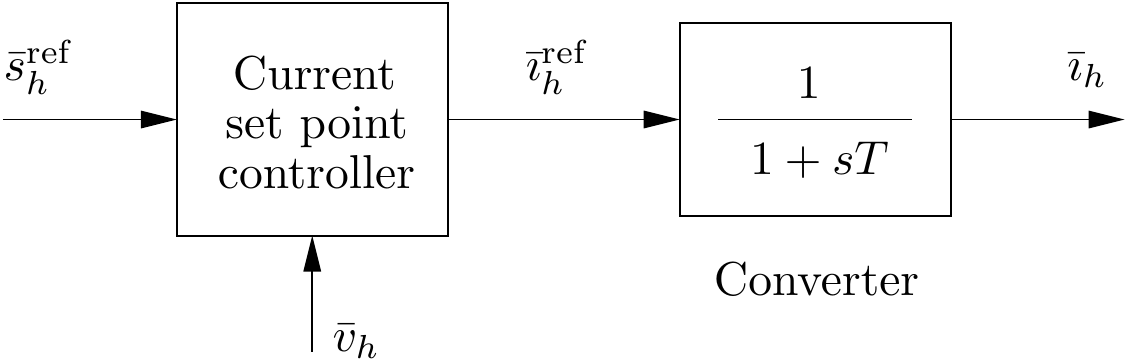}} \\
  \vspace{4mm}
  \resizebox{0.99\linewidth}{!}{\includegraphics{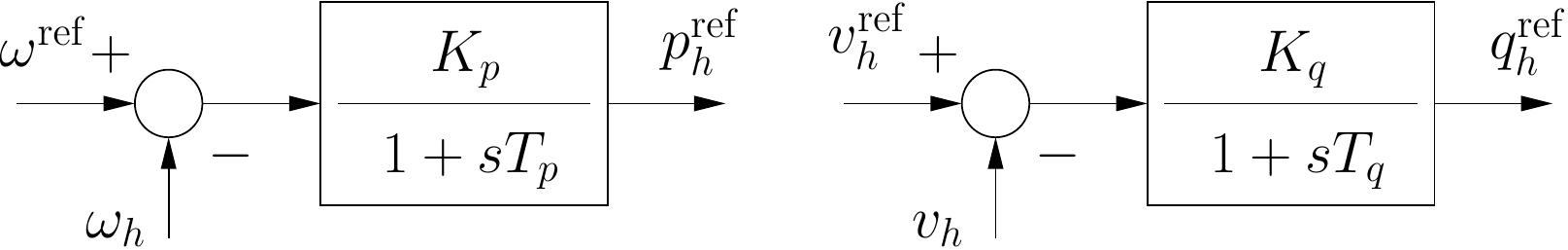}}
  \caption{Simplified scheme of a \ac{cig} and its controllers.}
  \label{fig:cig:model}
\end{figure}

Figure \ref{fig:syn} shows the imaginary part of the complex frequency
as obtained for the WSCC 9-bus system using a 4th and a 2nd order
model of the synchronous machines, as well as for a scenario where the
machine at bus 3 is substituted for a \ac{cig}.  The model of the
\ac{cig} is shown in Fig.~\ref{fig:cig:model}.  The results obtained
using \acp{pll} matches well the ``exact'' results obtained with
proposed method (indicated with $\rm CF$ in the legends) except for
the numerical spikes that following the load disconnection and a small
delay that is due to the control loop of the \ac{pll}.  The proposed
formula is also robust with respect to noise as shown in
Fig.~\ref{fig:syn:ex4}.

The events create spikes in the estimation of $\omega$ obtained with
the PLL.  The behavior of $\omega$ as estimated with the complex
frequency, on the other hand, depends on the dynamics of the device
connected to the bus where the frequency is estimated.  It is smooth
for the 2nd order synchronous machine and show a small jump for the
4th order synchronous machine.  The latter behavior is due to the
dependence of the frequency on the voltage magnitude, which is modeled
as an algebraic variable in \eqref{eq:dae}.  On the other hand, the
behavior of the $\omega$ estimated with the complex frequency at the
\ac{cig} bus shows a relatively fast transient after the load
disconnection (see Fig.~\ref{fig:syn:ex3}). This is due to the
dynamics of the currents of the converter.

Figure \ref{fig:syn} also shows the results obtained using the
\ac{fdf} proposed in \cite{Divider}.  The \ac{fdf} matches closely the
results of the proposed method when considering the simplified 2nd
order model of the machines but introduces some errors if the machine
model includes rotor flux dynamics or when considering the \ac{cig}.

\subsection{\ac{vdl}}
\label{sub:vdl}

The power consumption of a \ac{vdl} is:
\begin{equation}
  \label{eq:vdl1}
  \sh = \ph + j \qh = -p_o \vh^{\gamma_p} - \jj \, q_o \vh^{\gamma_q} \, ,
\end{equation}
then:
\begin{equation}
  \label{eq:vdl2}
  \begin{aligned}
    \shdot &= -p_o \, \gamma_p \, \vhdot \, \vh^{\gamma_p-1} -
    \jj \, q_o \, \gamma_q \, \vhdot \, \vh^{\gamma_q-1} \\
    &=  (\gamma_p \, \ph + \jj \, \gamma_q \, \qh ) \, \rhoh \, ,
  \end{aligned}
\end{equation}
where it has been assumed that the exponents $\gamma_p$ and $\gamma_q$
are constant and that $p_o$ and $q_o$ vary ``slowly'' with respect to
$\vh$.  If $\gamma_p = \gamma_q = \gamma$, then
$\shdot = \gamma \, \sh \, \rhoh$, which generalizes the results
obtained in Section \ref{sub:yconst}.

As an application, we utilize \eqref{eq:vdl2} to estimate the
parameters $\gamma_p$ and $\gamma_q$ of a \ac{vdl} using a similar
technique as the one proposed in \cite{Freqload}.  From
\eqref{eq:sdot} and \eqref{eq:vdl2}, one obtains:
\begin{equation}
  \label{eq:vdl3}
  (\gamma_p \, \ph + \jj \, \gamma_q \, \qh ) \, \rhoh =
  \sk [\shk (\etah + \etak^*)] \, ,
\end{equation}
and, splitting real and imaginary parts:
\begin{equation}
  \label{eq:vdl4}
  \begin{aligned}
    \gamma_p &= (\ph \, \rhoh)^{-1} \, \Re \left \{ \sk \shk (\etah + \etak^*) \right \} \, , \\
    \gamma_q &= (\qh \, \rhoh)^{-1} \, \Im \left \{ \sk \shk (\etah + \etak^*) \right \} \, , \\
  \end{aligned}
\end{equation}
where the right-hand sides can be determined based on measurements.
The fact that $\rhoh \rightarrow 0$ in steady-state can create
numerical issues, which can be solved, as discussed in
\cite{Freqload}, using finite differences over a period of time
$\Delta t$, namely $\etah \approx \Delta \eh/\Delta t$,
$\etak^* \approx \Delta \ek^*/\Delta t$, and
$\rhoh \approx \Delta \uh/\Delta t$, as follows:
\begin{equation}
  \label{eq:vdl5}
  \begin{aligned}
    \hat{\gamma}_p &\approx (\ph \, \Delta \uh)^{-1} \,
    \Re \left \{ \sk \shk (\Delta \eh + \Delta \ek^*) \right \} \, , \\
    \hat{\gamma}_q &\approx (\qh \, \Delta \uh)^{-1} \, \Im \left \{ \sk \shk
      (\Delta \eh + \Delta \ek^*) \right \} \, .
  \end{aligned}
\end{equation}
Equations \eqref{eq:vdl4} and \eqref{eq:vdl5} generalize the empirical
formulas to estimate $\gamma_p$ and $\gamma_q$ proposed in
\cite{Freqload}.  The latter, in fact, can be obtained from
\eqref{eq:vdl5} by approximating $\shk \approx - j \Bhk$.

\begin{figure}[htb!]
  \centering
  \begin{subfigure}[t]{\linewidth}
    \resizebox{0.49\linewidth}{!}{\includegraphics{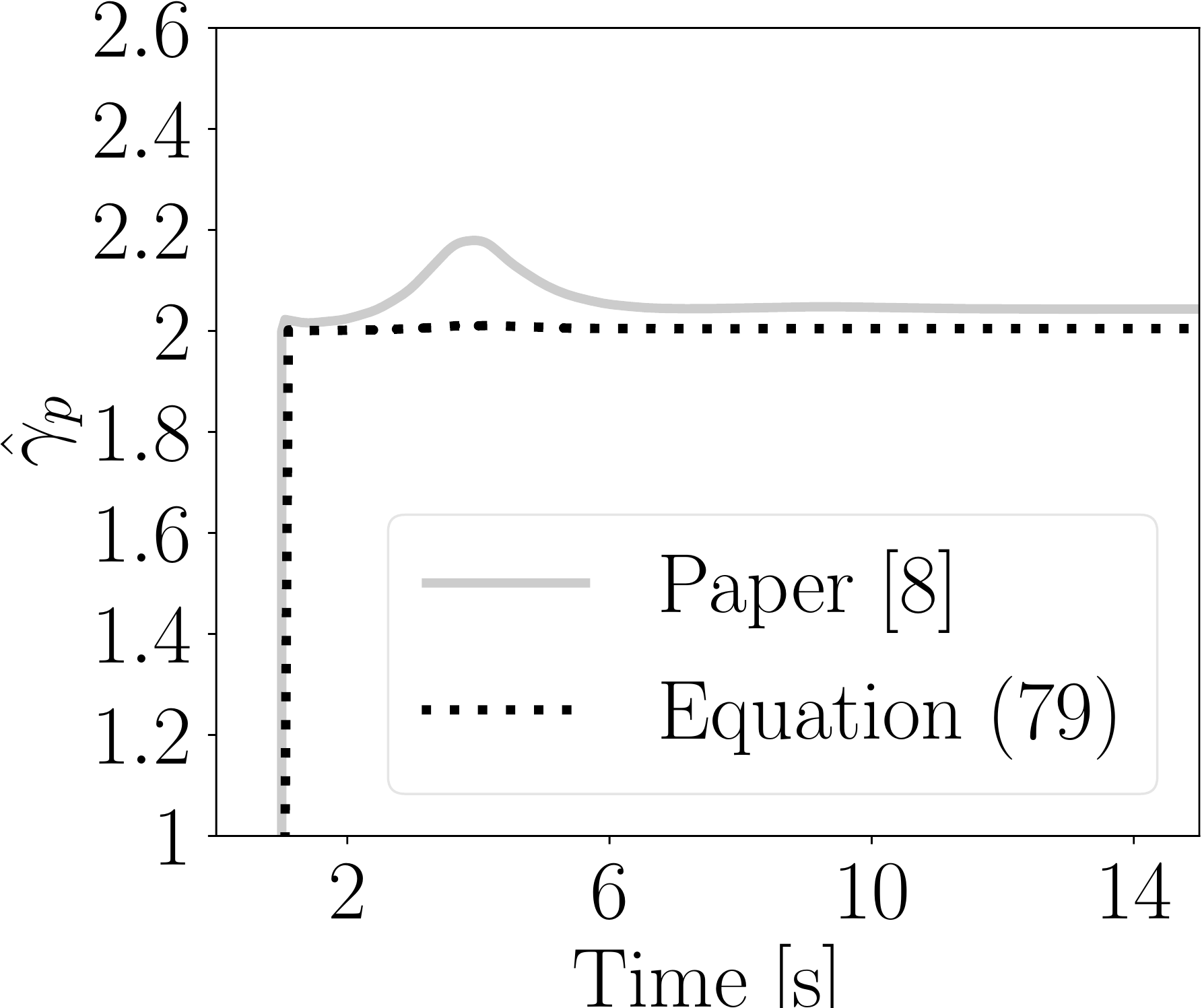}}
    \resizebox{0.49\linewidth}{!}{\includegraphics{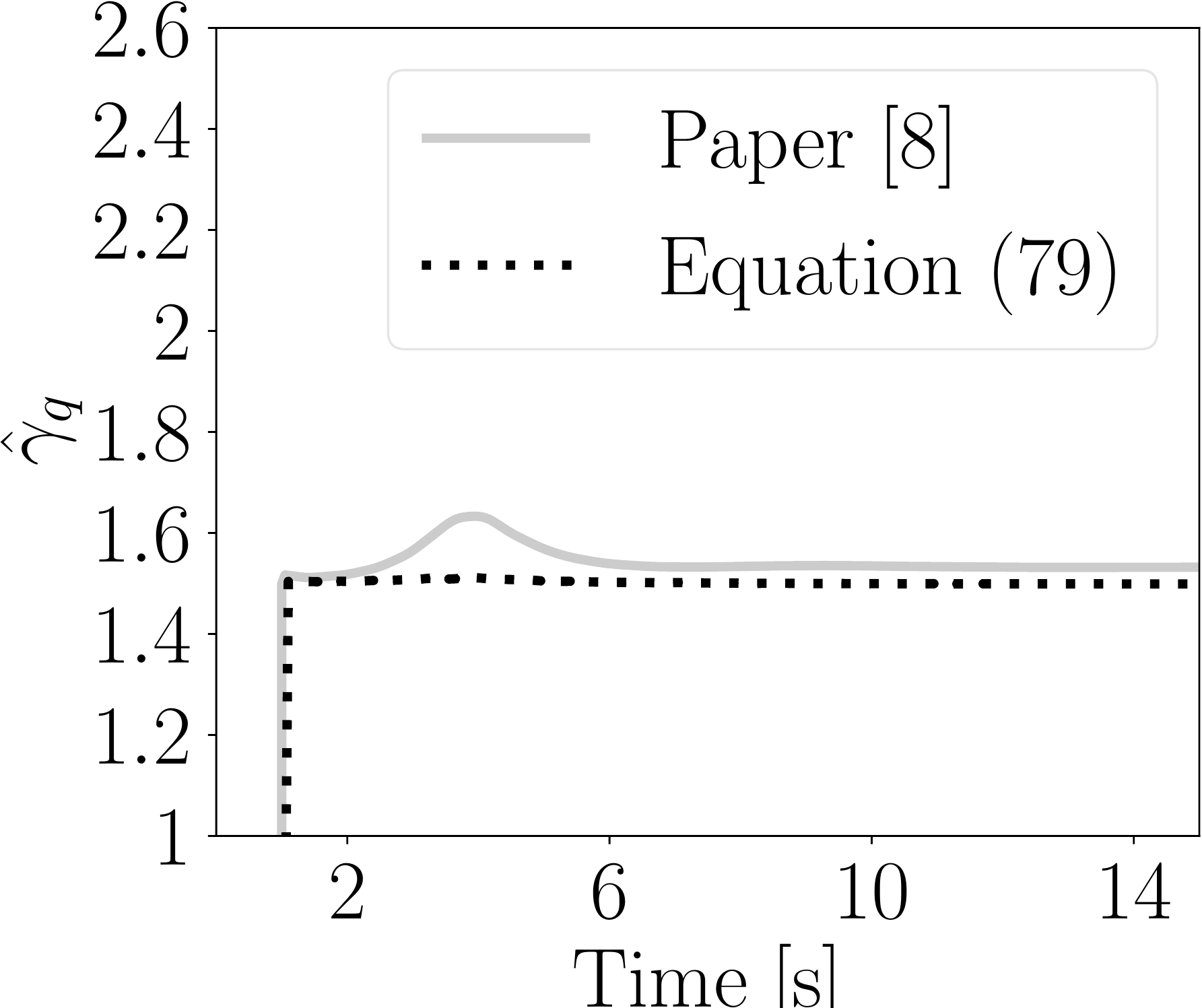}}
    \caption{Ratio of transmission lines: $R/X \ll 1$.}
    \label{fig:vdl:ex1}
  \end{subfigure} \\ \vspace{3mm}
  \begin{subfigure}[t]{\linewidth}
    \centering
    \resizebox{0.49\linewidth}{!}{\includegraphics{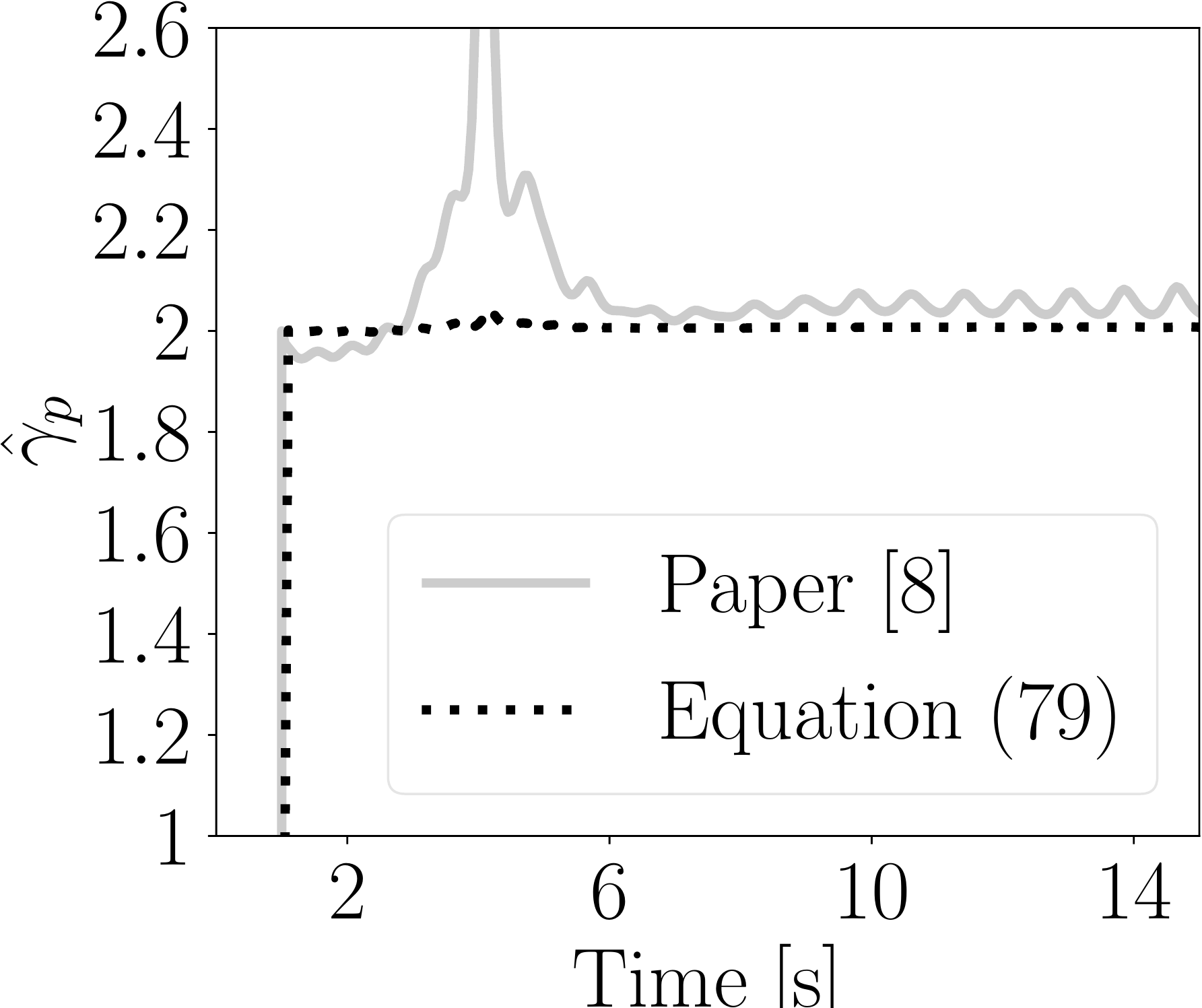}}
    \resizebox{0.49\linewidth}{!}{\includegraphics{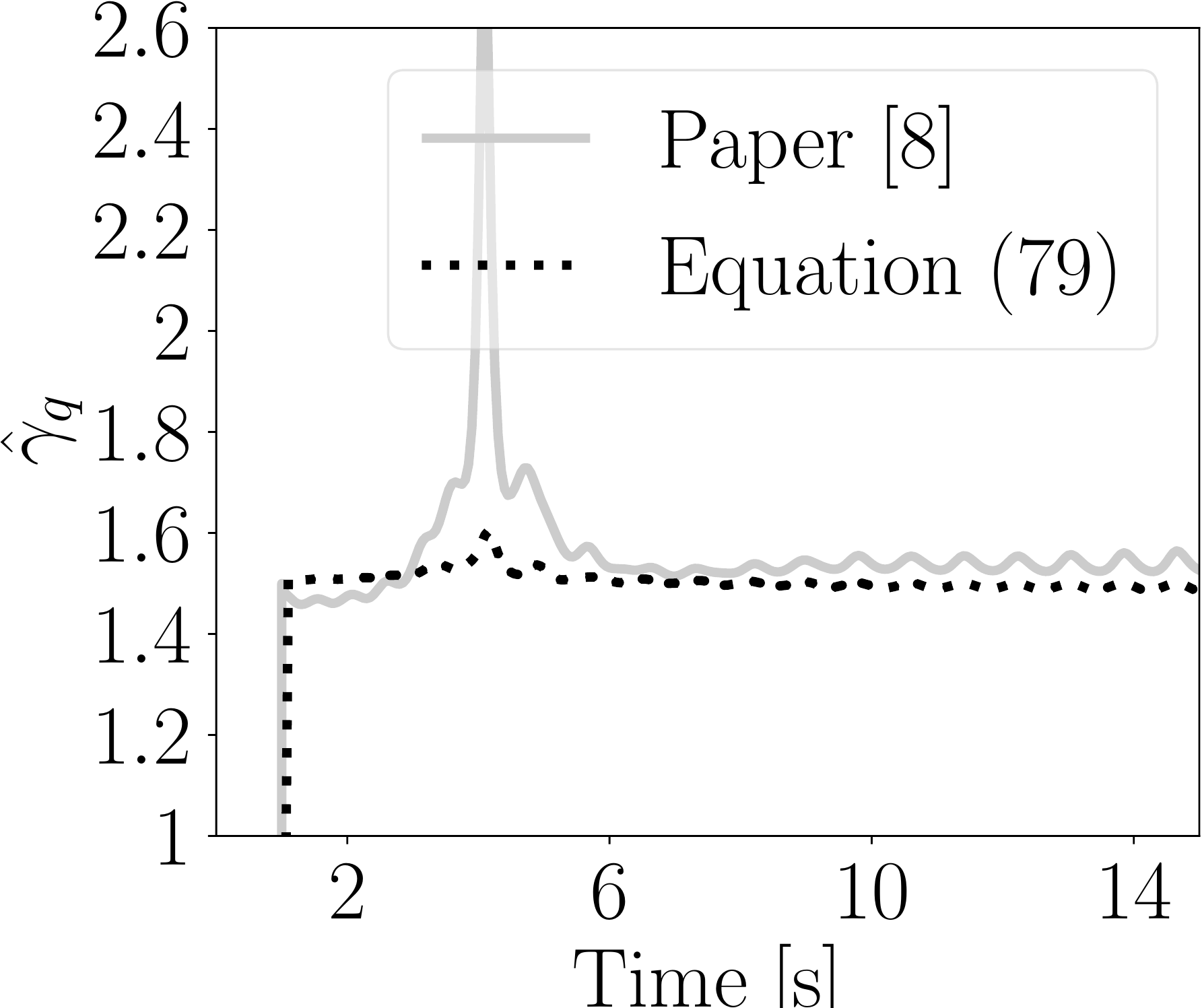}}
    \caption{Ratio of transmission lines: $R/X \approx 1$.}
    \label{fig:vdl:ex2}
  \end{subfigure}
  \caption{Estimation of the exponents of the \ac{vdl} connected at
    bus 8 following the disconnection of 15\% of the load at bus 5 at
    $t=1$ s for the WSCC 9-bus system.}
  \label{fig:vdl}
  \vspace{2.5mm}
\end{figure}

Figure \ref{fig:vdl} shows the results obtained for the WSCC 9-bus
system where the bus connected at bus 8 is a \ac{vdl} with
$\gamma_p = 2$ and $\gamma_q = 1.5$.  The results show that
\eqref{eq:vdl5} is, as expected, more precise than the approximated
estimations based on the expression proposed in \cite{Freqload}.
Equation \eqref{eq:vdl5} is, in fact, an exact expression and its
accuracy depends exclusively on the accuracy of the measurements of
$\eh$ and $\ek$, which can be obtained, for example, with PMUs.  If
the $R/X$ ratio of the transmission lines of the system is changed to
resemble that of a distribution system, \eqref{eq:vdl5} appears also
numerically more robust than its approximated counterparts (see
Fig.~\ref{fig:vdl:ex2}).

\subsection{Converter-Interfaced Generation}
\label{sub:cig}

\begin{figure}[htb!]
  \centering
  \resizebox{\linewidth}{!}{\includegraphics{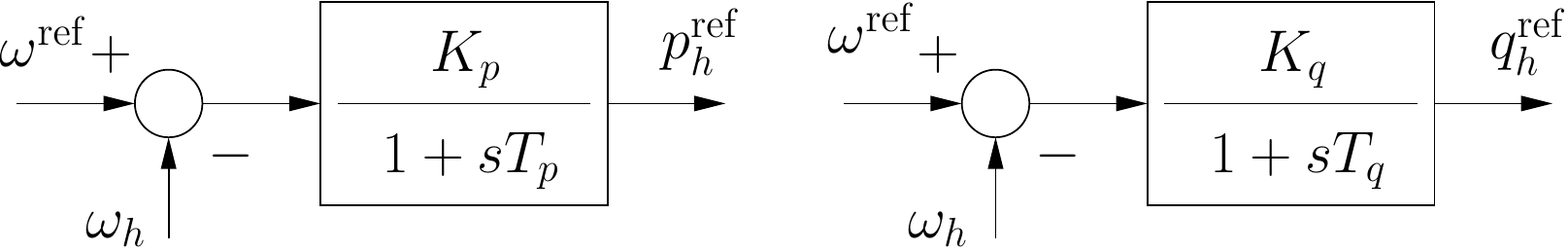}}
  \caption{Alternative frequency control through active and reactive
    powers for the \ac{cig} of Fig.~\ref{fig:cig:model}.}
  \label{fig:cig}
  \vspace{2.5mm}
\end{figure}

This last example illustrates an application of the approximated
expression \eqref{eq:pdot:app2}.  We focus in particular on the link
between $\wvec$ and $\dot{\bfg q}$ through the resistances of network
branches.  Using again the WSCC 9-bus system and substituting two
synchronous machines with \acp{cig}, we compare the dynamic response
of the system following a load variation using the control scheme of
Fig.~\ref{fig:cig:model} (Control 1), and the control scheme shown in
Fig.~\ref{fig:cig} (Control 2).  The latter regulates the frequency by
both the active and reactive powers of the \acp{cig}.

\begin{figure}[htb!]
  \centering
  \resizebox{0.98\linewidth}{!}{\includegraphics{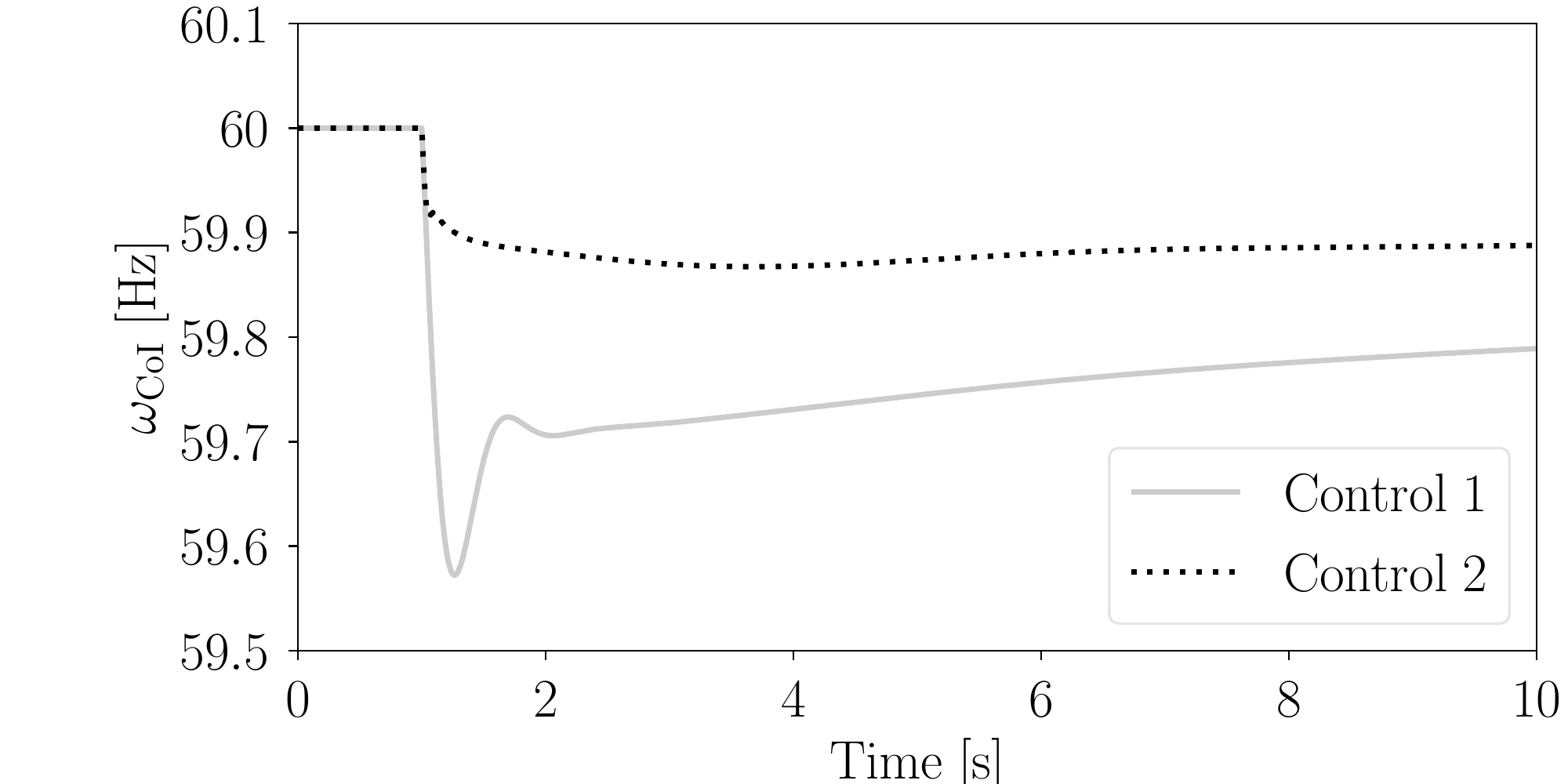}} \\
  \vspace{2mm}
  \resizebox{0.98\linewidth}{!}{\includegraphics{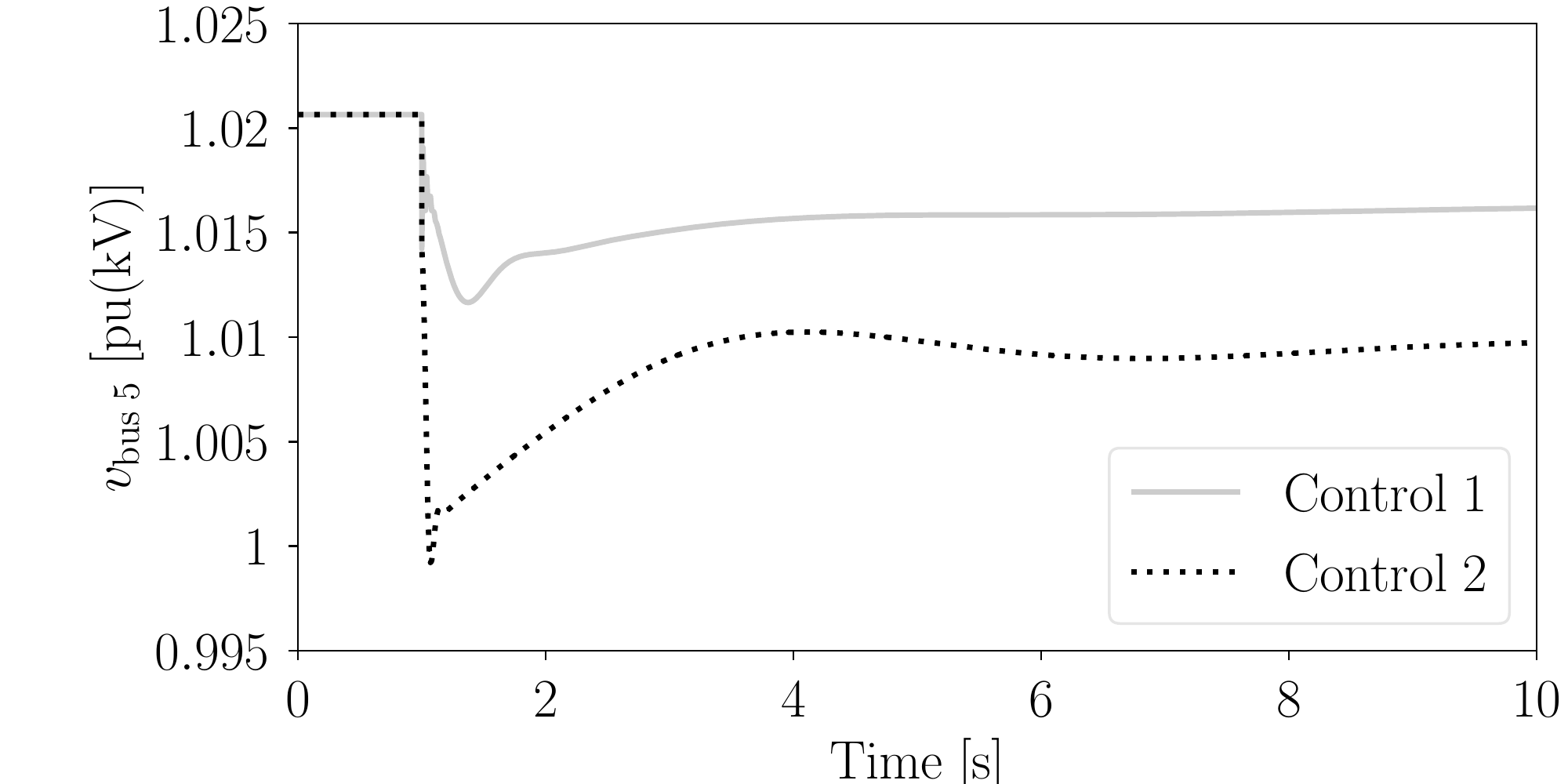}}
  \caption{Frequency of the \ac{coi} and voltage at bus 5 following
    the connection of $p= 0.25$ pu at bus 5 at $t=1$ s for the WSCC
    9-bus system with high penetration of \ac{cig}.}
  \label{fig:der}
\end{figure}

Figure \ref{fig:der} shows that Control 2 is more effective than
Control 1 to reduce the variations of the frequency.  This result
indicates that the relationship between $\bfg q$ and $\bfg \wvec$ is
not weak and can be exploited to improve the frequency response of
low-inertia systems.  This result is predicted by \eqref{eq:pdot:app2}
and hence by \eqref{eq:sdot}.

The simplified converter-interfaced generator model shown in
Fig.~\ref{fig:cig}, with differential equations:
\begin{equation}
  \label{eq:cig1}
  \begin{aligned}
    T_p \, \dot{p}_h &= K_p (\omega^{\rm ref} - \wh) - p_h \, , \\
    T_q \, \dot{q}_h &= K_q (v^{\rm ref} - \vh) - q_h \, .
  \end{aligned}
\end{equation}

% If $T_p = T_q = T$ and $K_p = K_q = K$, then:
% %
% \begin{equation}
%   \label{eq:cig2}
%   T \, \shdot = K \, [ (\omega^{\rm ref} - \wh) + j (v^{\rm ref} - \vh) ] - \sh \, ,
% \end{equation}
% %
% \begin{equation}
%   \label{eq:cig3}
%   \begin{aligned}
%     T \, \shdot &= K \, [ (\omega^{\rm ref} - \wh) + j (\rhoh^{\rm ref} - \rhoh) ] - \sh  \\
%     &= K \, ( \etah^{\rm ref} - \etah ) - \sh \, ,
%   \end{aligned}
% \end{equation}

\section{Conclusions}
\label{sec:conclusion}

The paper introduces a new physical quantity, namely, the
\textit{complex frequency}.  This quantity allows writing the
differential of power flow equations in terms of complex powers and
voltages at network buses.  The most significant property of the newly
defined complex frequency is its ability to give a more robust and
clean indication of frequency than what generally accepted in the
literature, especially to describe the behavior of the frequency at
buses close to a disturbance.  For example, it is well known that many
other methods result in meaningless spikes in frequency at the
inception and clearing of faults and other sudden disturbances.  The
proposed definition provides a solution to this issue.

The proposed definition of complex frequency is in accordance with the
commonly accepted definition of frequency and generalizes it.  In
fact, if the magnitude of the voltage is constant, then $\rho = 0$ and
the magnitude of the complex frequency coincides with
\eqref{eq:IEEE118}.  The common definition of frequency is thus a
special case of the complex frequency.  On the other hand, if the
magnitude of the voltage is not constant, then, the complex frequency
extends the concept of frequency by including an additional term,
$\rho$.

Then, the paper shows how the complex frequencies are related to the
time derivatives of the voltages and to the power/current injections
at the network buses.  In this vein, noteworthy expressions are
\eqref{eq:sdot}, \eqref{eq:sdot:alt2} and \eqref{eq:idot}.  It is also
shown that these expressions are a generalization of the \ac{fdf}
proposed in \cite{Divider}.

The proposed framework can be used, for example, for evaluating how
good is the estimation of the frequency obtained with a PLL and also
appears as an useful tool to design novel and efficient controllers.
The several analytical and numerical examples discussed in Sections
\ref{sec:derivation} and \ref{sec:modeling} show the prospective
applications of the proposed theoretical approach.  The newly
introduced term $\rho$ can be also expected to have practical
applications.  An example is given in Section \ref{sub:vdl}.

Future work will focus on further elaborating the expression
\eqref{eq:sdot} and exploiting its features for the control and state
estimation of power systems.  A relevant question worth of further
study is how to take into account fast dynamics, such as transmission
line dynamics, in the evaluation of the complex frequency.  It appears
relevant also to combine the proposed complex frequency approach to
some of the techniques described in \cite{Paolone:2020}.  Particularly
interesting, for example, is the ``beyond-phasor'' approach based on
the Hilbert transform proposed in \cite{8928939}.

\appendices

\section{Time-Derivative of the Voltage}
\label{app:vdot}

This appendix provides the proof of \eqref{eq:vdot}.  With this aim,
let us consider the $h$-th element of \eqref{eq:vdot}, namely
$\vhbar = \vh \angle \th = \vh (\cos \th + \jj \sin \th)$.  The time
derivative of $\vhbar$ with respect to the $\rm dq$-axis reference
frame gives:
\begin{align*}
  \vhbardot &= \vhdot \angle \th + \jj \, \vh \, \wh \angle \th \\
            &= \vh \left ( \rhoh \angle \th + \jj \, \wh \angle \th \right ) \\
            &= \vh \angle \th \left ( \rhoh + \jj \, \wh \right ) \\
            &= \vhbar \, \etah \, ,
\end{align*}
where the following identities hold:
\begin{align*}
  \frac{d}{dt} \angle \th &= \wh \, (- \sin \th + \jj \cos \th) \\
  &= \jj \, \wh \, (\cos \th + \jj \sin \th) \\
  &= \jj \, \wh \, \angle \th \, .
\end{align*}

\section{Complex Frequency of Current Injections}
\label{app:idot}

% It appears interesting to note that, in general and in transient
% conditions, the complex frequency of the current injections at
% network buses is not the same as the complex frequency of the
% voltages at the same buses.

Let us assume that $\ih = \ii_h \angle \beta_h$, then, one has:
\begin{equation}
  \label{eq:ihdot}
  \begin{aligned}
    \ihdot = \ih \left ( \frac{\dot{\ii}_h}{\ii_h} +
      \jj \dot{\beta}_h \right ) = \ih \, \xih \, ,
  \end{aligned}
\end{equation}
where $\xih$ is the complex frequency of the current injection at bus
$h$ which is determined using the same procedure that leads to
\eqref{eq:vdot} and that is described in the Appendix.

%  It is worth noticing that, in general, $\xih \ne \etah$.
% This is due the dynamic behavior of the device(s) connected at bus $h$
% and that define the relationship between voltage and current.  If this
% relationship is nonlinear, the current is not a perfect sine wave even
% if the voltage is.  For example, the saturation of the iron core of
% transformers is a common source of distortions in the wave of the
% current.  These distortions can be interpreted in terms of harmonics
% (and thus the fundamental frequency of the current will be the same as
% the one of the voltage) or assuming
% $\imath_h(t) = I_h(t) \sin_h(\beta(t))$ with time-varying $I_h$ and
% $\dot{\beta}_h$ even if $v_h(t) = V_h \sin(\wo t)$ with constant $V_h$
% and $\wo$ (thus leading to $\xih \ne \etah$).

A relevant special case is that of constant power factor devices
(either generators or loads), for which
$\beta_h = \th + \varphi_{ho}$, where $\varphi_{ho}$ is the constant
power factor angle. Then:
\begin{equation}
  \label{eq:psidot}
  \dot{\beta}_h = \dot{\theta}_h + \dot{\varphi}_{ho} = \wh \, .
\end{equation}
On the other hand, $\Re \{ \bar{\xi}_h \} = \varrho_h$ only if
$\ii_h = k \vh$.  Since a constant impedance has constant power factor
and its current is proportional to the voltage, it satisfies the
condition $\xih = \etah$, $\forall t$.

\section*{Acknowledgments}

The author wishes to thank Dr.~Ioannis Dassios, University College
Dublin, for carefully revising the mathematical derivations given in
the paper.

%======================================================================
%\bibliographystyle{IEEEtran}
%\bibliography{refs}

%======================================================================

%======================================================================
\begin{biography}
  [{\includegraphics[width=1in,height=1.25in,clip,keepaspectratio]{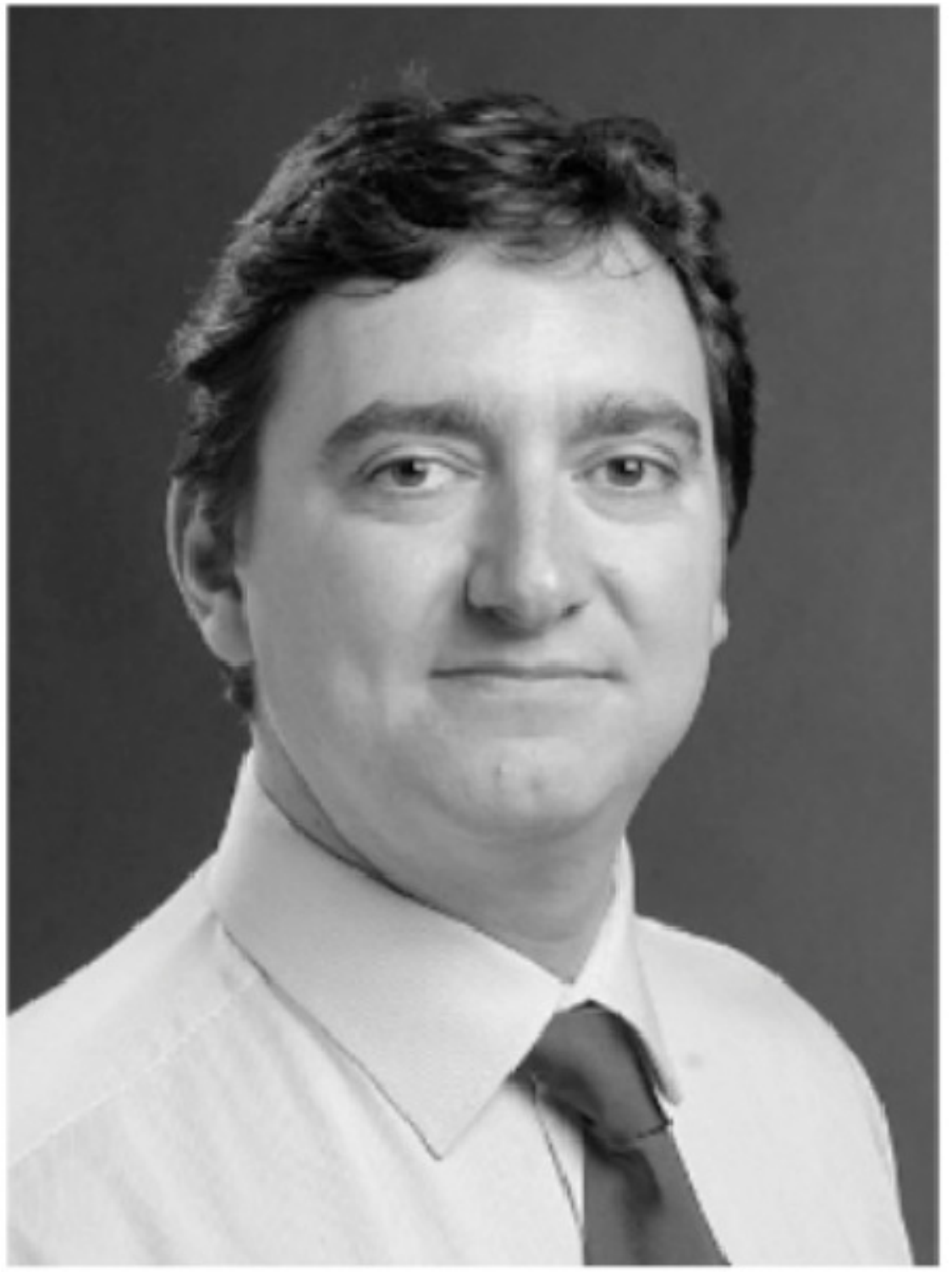}}]
  {Federico Milano} (F'16) received from the University of Genoa,
  Italy, the ME and Ph.D.~in Electrical Engineering in 1999 and 2003,
  respectively.  From 2001 to 2002, he was with the Univ.~of Waterloo,
  Canada.  From 2003 to 2013, he was with the Univ.~of Castilla-La
  Mancha, Spain.  In 2013, he joined the Univ.~College Dublin,
  Ireland, where he is currently Professor of Power Systems Control
  and Protections and Head of Electrical Engineering.  His research
  interests include power systems modeling, control and stability
  analysis.
\end{biography}

\vfill

\end{document}